\documentstyle[aps,eqsecnum,epsf,cite]{revtex}

\newcommand{\bq}{{\bbox q}}
\newcommand{\bw}{{\bbox w}}
\newcommand{\tbw}{\bbox{\widetilde{w}}}
\newcommand{\bh}{{\bbox h}}

\begin{document}
\draft
\title{Spectrum of Cosmological Perturbations in the 
One-Bubble Open Universe} 
\author{Jaume Garriga$^1$, Xavier Montes$^1$, Misao
  Sasaki$^2$ and Takahiro Tanaka$^2$} 
\address{${}^1$IFAE, Edifici C,
  Universitat Aut{\`o}noma de Barcelona, E-08193 Bellaterra, Spain\newline
  ${}^2$Department of Earth and Space Science, Graduate School of Science,
  Osaka University, Toyonaka 560, Japan}
 \maketitle

\medskip
\centerline{\bf\today}%{2 Jun 1998}

\begin{abstract}
The spectrum of cosmological perturbations in the context of the
one-bubble open inflation model is discussed, taking into account
fluctuations of the metric. We find that, quite generically,thin wall
single field models have no supercurvature modes. However, single field
models with supercurvature modes do exist. In these models the density
parameter $\Omega$ becomes a random variable taking a range of values
inside of each bubble. We also show that the model dependence of the
continuous spectrum for both scalar and tensor-type perturbations is
small as long as the kinetic energy density of the background field does
not dominate the total energy density. We conclude that the spectrum of
the density perturbation predicted in the single-field model of the
one-bubble open inflation is rather robust. We also consider the
spectrum of scalar and tensor perturbations in a model of the
Hawking-Turok type, without a false vacuum. 
\end{abstract}

\section{Introduction}

In the standard inflationary universe scenario, both the flatness and
the homogeneity problems are simultaneously solved by the same
mechanism, i.e., the accelerated expansion of the cosmic length scale.
Therefore, if we want to solve the homogeneity problem, the flatness
problem is automatically solved. This would make it impossible to
create an open universe. Recently, however, attention has focused on a
scenario which solves this difficulty. It is called the one-bubble
open inflation scenario. The basic idea was first proposed by Gott III
\cite{Gott,ColDeL}.  There are two main classes of models that realize
Gott's idea: the single-field model, which was developed in
\cite{models,turok}, and the two-field models proposed in
\cite{modelt}.  Here we shall basically focus on the single-field
model, but most of the results presented in this paper can be applied
to models with two fields.

In the single-field model we assume a scalar field with a potential
such as the one shown in Fig.~1.  Initially, the field is trapped in
the false vacuum, where the vacuum energy drives the exponential
expansion of the universe. This exponential expansion solves the
homogeneity problem. After a sufficiently long period of inflation,
bubble nucleation occurs through quantum tunneling.  This process is
described by the $O(4)$-symmetric bounce solution \cite{ColDeL}, a
solution of the Euclideanised equation of motion that connects the
initial and final configurations. These configurations represent the
state before and after tunneling, respectively.  In the lowest WKB
order, the classical evolution after tunneling is determined by the
analytic continuation of the bounce solution.  After the analytic
continuation, the $O(4)$-symmetry changes into the $O(3,1)$-symmetry,
which is just the symmetry of an open Friedmann-Robertson-Walker
universe. Thus, in the lowest order approximation, we obtain a model
that explains the creation of an open and homogeneous universe.
However, at this stage the universe is almost empty and the cosmic
expansion is dominated by the curvature term in the Friedmann
equation. Hence, as long as only ordinary matter fields are assumed,
the universe would stay curvature dominated forever.  Therefore, in
order to solve the entropy problem, a second stage of inflation inside
the nucleated bubble is needed.

\begin{figure}[htb]
\centerline{\epsfbox{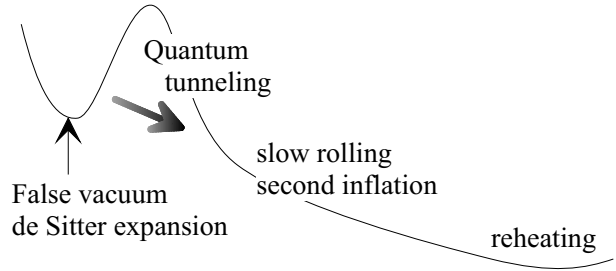}}
\caption{A schematic picture of the inflaton potential 
for a single-field model.} 
\end{figure}

\begin{figure}[htb]
\centerline{\epsfbox{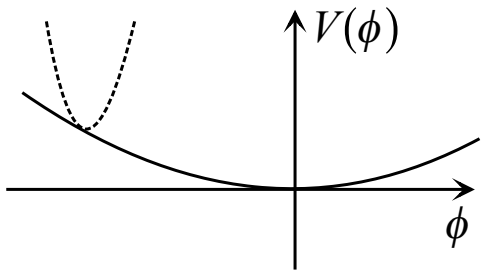}}
\caption{
A schematic picture of the potential of the second inflaton field for 
a two-field model.} 
\end{figure}

Although this model is simple, we need to assume a rather unusual form
of the inflaton potential (see however \cite{Lindetoy}).  To solve
this problem of naturalness, models with two fields were proposed,
where two different inflatons would drive the first and second stages
of inflation. The first inflaton field, $\sigma$, has a double well
potential. When the $\sigma$-field is in the false vacuum, the energy
density of the universe is dominated by the vacuum energy of the
$\sigma$-field.  At this stage, the second inflaton field, $\phi$, has
the potential shown by the dashed line in Fig.~2. After a sufficiently
long period of inflation the $\sigma$-field makes a phase transition
to the true vacuum through bubble nucleation.  It is assumed that due
to the coupling to the $\sigma$ field, the potential of the
$\phi$-field changes into the one shown by the solid line in Fig.~2.
Then, the slow roll of $\phi$ from the old minimum to the new one
drives the second period of inflation in the nucleated bubble.

In some cases, the instanton describing tunneling has an approximate
zero mode which can shift back and forth the value where the slow roll
field lands after tunneling.  In those cases, the density parameter in
the resulting universe becomes a random variable, and one has an
ensemble of large regions inside of each bubble where the density
parameter $\Omega$ (measured at a fixed temperature) takes a range of
values. This can be understood as follows.

The existence of an approximate zero mode of the instanton implies
that, in the real time evolution of the bubble, there is a
supercurvature mode (whose wavelength exceeds the curvature scale).
Quantum excitations of this mode can change the average value of the
slow roll field inside the bubble on scales larger than the Hubble
radius. As a result, the amount of slow roll inflation in different
regions will be different, leading to a range of values of $\Omega$
\cite{anth}.  In particular, even if the instanton leads to a
non-inflating value of the field after tunneling, localized
fluctuations can create inflating islands which locally resemble open
universes.  This scenario is called ``quasi-open'' inflation
\cite{modelq,anth}.  Quasi-open inflation was originally discussed in
the context of two-field models, but as we shall see, models with a
single field can also have supercurvature modes with similar
consequences.

Finally, Hawking and Turok \cite{HT} have recently proposed that it is
possible to create an open universe from nothing in a model without a
false vacuum.  The instanton describing this process is singular, and
therefore its validity has been subject to question \cite{htall}.
Nevertheless, it has also been pointed out that the quantization of
linearized perturbations in the singular background is well posed
\cite{jaume,cohn}. Therefore, provided that one can make sense of the
instanton by appealing to an underlying theory where the singularity
is resolved, it appears that the details of that theory need not be
known in order to calculate the spectrum of cosmological
perturbations.

\begin{figure}[htb]
\centerline{\epsfbox{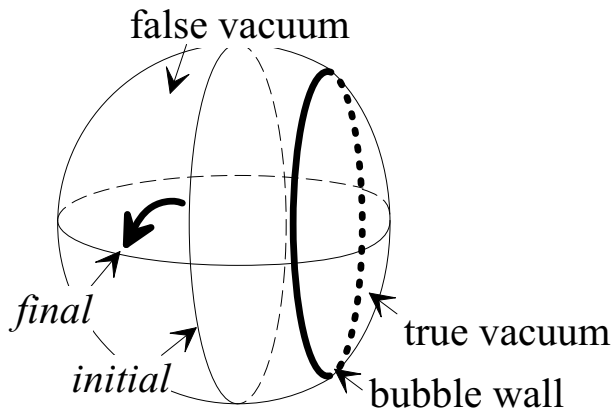}}
\caption{
The Coleman-De~Luccia instanton for the case that the spacetime
dimension is two, embedded in the three dimensional Euclidean space. 
}
\end{figure}

\begin{figure}[htb]
\centerline{\epsfbox{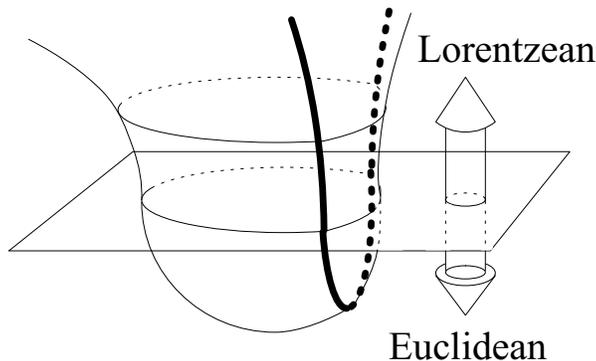}}
\caption{
  The analytic continuation of the Coleman-De~Luccia instanton to the
  Lorentzian region. The lower half is the same as that shown in Fig.3
  and represents the Euclidean region. The upper half represents the
  continuation to the Lorentzian region. }
\end{figure}

In order to construct an acceptable cosmological model, we must also
show that the expected spectrum of primordial fluctuations is
compatible with observations. Since there is an initial de Sitter
expansion phase, it is natural to assume that the state before
tunneling is the so-called Bunch-Davies vacuum state.  Then the
question that we must answer is the following: ``What quantum state
results after $O(4)$-symmetric bubble nucleation starting from the
Euclidean vacuum state?'' In the lowest order WKB approximation, the
tunneling process is described by the Coleman-De~Luccia bounce
solution \cite{ColDeL}. In order to extract information about the
quantum fluctuations, it is necessary to develop the WKB approximation
to the next order \cite{tunnel}.  The resulting prescription to find
the quantum state after tunneling can be summarized as follows.

First of all, since we shall deal with the metric perturbations 
which contain gauge degrees of freedom,  
we need to find the reduced action that contains only 
the physical degrees of freedom. We denote one of these degrees of freedom
by $\varphi(x)$. With an appropriate choice of the variable $\varphi(x)$, 
the action can be reduced to that of the 
massive scalar field with variable mass, 
\begin{equation}
 S_{\varphi}=
\int\biggl(-{1\over2}g^{\mu\nu}\partial_{\mu}\varphi 
 \partial_{\nu}\varphi-{1\over 2}{\cal M}^2(\eta_C)\varphi^2 
 \biggr)\sqrt{-g}d^4x. 
\label{action}
\end{equation}
{}From this reduced action, one can calculate expectation values
following standard methods. First, we need to find a complete set 
of functions $v_{\bf k}$ which obey the field equation, 
\begin{equation}
  \Bigl[\nabla^\mu\nabla_\mu -{\cal M}^2(\eta_C)\Bigr]
v_{\bf k}(\eta,{\bf x})=0,
\label{KGeq}
\end{equation}
in the Lorentzian region, and which are {\em regular on the lower
  hemisphere} shown in Fig.~4. This restriction is due to the choice
of the Bunch-Davies vacuum state before tunneling. However, the same
restriction arises in the no-boundary proposal for the wave-function
of the Universe \cite{NB}, if Fig. 4 is interpreted as describing the
creation from nothing of a universe containing a bubble. The same
restriction will therefore be used in the case of the Hawking-Turok
model discussed in Appendix D.  Here, it is important to note that ``a
complete set'' means a set of all modes which can be Klein-Gordon
normalized on a Cauchy surface $\Sigma$ of spacetime,
\begin{eqnarray}
  \left\langle v_{{\bf k}}\,,v_{{\bf k}'}\right\rangle
:=-i \int_{\Sigma}d\Sigma_\mu g^{\mu\nu} 
  \left\{v_{\bf k} \partial_\nu {\overline{v_{{\bf k}'}}}
  -(\partial_\nu v_{{\bf k}})\overline{v_{{\bf k}'}}\right\}
  =\delta_{{\bf k}{\bf k}'}.
\end{eqnarray}
Once they are obtained,
the quantum fluctuations of the field are described
by the ``vacuum state", $|\Psi\rangle$, that satisfies 
$\hat a_{\bf k}|\Psi\rangle=0$
for any ${\bf k}$ where the annihilation operator $\hat a_{\bf k}$ 
is defined by the decomposition of the perturbation field:
\begin{equation}
  \hat\varphi=\sum_{\bf k} (v_{\bf k} \hat a_{\bf k} + 
 {\overline{v_{\bf k}}}\hat a_{\bf k}^\dagger).
\end{equation}
Thus the mode functions $v_{\bf k}$ play the role of
positive frequency functions.

The most tedious part in the above prescription is to find 
the reduced action. This is due to the fact that the time-constant 
hypersurface that reflects the maximal symmetry of background solution 
is not a Cauchy surface, and hence it is not appropriate 
to normalize modes. Therefore, we need to perform the reduction 
of the action on spacelike hypersurfaces which cut right through 
the bubble, and which are therefore not homogeneous. 
This renders the decomposition into scalar, vector and tensor 
modes into a rather unfamiliar form. 
In the end, however, the three standard physical degrees of 
freedom are identified \cite{prev}. 
When they are analytically continued to the open universe, 
one becomes the usual scalar-type perturbation and the other 
two becomes the even and odd parity tensor-type  
perturbations. 

Along the lines of the above prescription, the prediction of the power
spectrum of perturbations was given by many authors
\cite{spec,YST,STY}.  Until recently, however, the reduced action
including the metric perturbations was not known, and for the
scalar-type perturbation the studies were limited to the case in which
the metric perturbation was neglected. Clearly, this should be a
reasonable approximation in the weak gravity limit. Nevertheless, it
turns out that the correspondence between the case where metric
perturbations are neglected and the case where they are included is
rather non-trivial.

In \cite{prev} we found that, at least formally, the metric
perturbation has a dramatic effect on the spectrum for the scalar
modes, even in the limit of weak gravitational coupling. When metric
perturbations are ignored, there is a discrete wall fluctuation mode
which gives a finite contribution to observables. However, when we
incorporate the metric perturbation, this mode disappears.
Fortunately, the contribution from the disappeared scalar-type
perturbation reappears in the low frequency spectrum of even-parity
tensor modes \cite{TStensor,STY}.  In particular, we found that in the
weak gravity limit the change of assignment of the wall fluctuation
modes from scalar to tensor does not affect the observational
quantities, such as the multipole moments of the temperature
fluctuation of the cosmic microwave background. Note that the wall
fluctuation mode is the Goldstone mode associated with the breakdown of
spatial symmetry, and we find that it is ``eaten up'' by the
gravitational degrees of freedom.  This is reminiscent of the fate of
Goldstone bosons in gauge theories.

In the present paper we shall continue the work of \cite{prev},
investigating the full power spectrum of scalar perturbations. In
particular, we shall consider the question of the existence of
supercurvature scalar modes\cite{YST,lythw}. In the previous studies in
which the metric perturbation was neglected it was found that, quite
generically, there are no supercurvature modes for thin wall
single-field models (other than the wall fluctuation mode). It was
also shown that the contribution to the CMB anisotropy due to the
continuous spectrum is almost independent of the details of the
potential barrier. However, it is still uncertain whether such a
robust prediction of the spectrum stays correct or not when we
incorporate the effect of the metric perturbation. Here we shall
clarify this point. As for the tensor-type perturbations, the studies
in which the perturbation of the scalar field was neglected was done
in \cite{TStensor,buchercohn}. Fortunately, this was found not to
alter the resulting spectrum at all. However, the effect of the
evolution after nucleation of a bubble has not been studied so far.

In previous studies, a simplified model was used
to describe the background geometry inside the bubble, assuming 
that it was a pure de Sitter space.
In this paper, we extend our analysis also taking into account 
that both the background geometry and the scalar field configuration 
are non-trivial.

The paper is organized as follows.  In Section II we first describe the
lowest WKB order picture of the bubble nucleation.  After that, we
briefly recall the resulting reduced action for the perturbation on
this background obtained in our previous paper\cite{prev}.  In Section
III we consider the question of existence of scalar supercurvature
modes. For this purpose, several new techniques are introduced.  In
particular we show that the spectra in the weak gravity limit and in
the case without gravity can be related to each other using the
formalism of supersymmetric quantum mechanics. We also introduce a
method for reconstructing the inflaton potential barrier starting from
a given bubble profile for the tunneling field. This is used to find
examples which do posess supercurvature modes. In particular, we
discuss an generalization of the flat space Fubini instanton
\cite{fubini} to de Sitter space. This example is useful in order to
illustrate the role played by the scalar supercurvature modes. In
Sections IV and V we consider the spectrum of the scalar-type and
tensor-type perturbations, respectively.  Section VI is devoted to the
conclusions.

A number of issues are discussed in the appendices. In Appendix A we
perform the canonical quantization of cosmological perturbations using
the Dirac formalism. We recover the results of our earlier work
\cite{prev}, where the quantization was performed using the
Fadeev-Jackiw reduction method.  In Appendix B we derive some bounds
on the ``superpotential'' which appears in the Schr\"odinger
fluctuation operator for scalar modes. In Appendix C, we consider a
class of models for which the evolution of perturbations can be solved
exactly. Finally, in Appendix D we find the spectrum of cosmological
perturbations in a soluble model of the Hawking-Turok type. We use the
notation, $\kappa=8\pi G$, and adopt the units $c=\hbar=1$.

\section{reduced action for perturbations}

We consider the system that consists of a minimally coupled 
single scalar field, $\Phi$, coupled with the Einstein gravity. 
We denote the potential of the field $\Phi$ by $V(\Phi)$.
In order to describe the $O(4)$-symmetric bounce solution, 
we consider a Euclidean configuration,
\begin{eqnarray}
 && ds^2=a(\eta_E)^2\left[d\eta_E^2+
 d\chi_E^2+\sin^2\chi_E(d\theta^2+\sin^2\theta d\varphi^2)\right],
\cr
 && \Phi=\phi(\eta_E). 
\label{metriceuclidean}
\end{eqnarray}
Then, the equation of motion for the background quantities becomes 
\begin{eqnarray}
 && \phi''+2{\cal H}\phi'-a^2 {\partial V}=0,
\label{phipp}\\
 && {\cal H}^2-1={\kappa\over 3}\left({1\over 2}{\phi'}^2
    -a^2V(\phi)\right), 
\label{calHH}\\
 && {\cal H}'-{\cal H}^2 +1 = -{\kappa\over 2}{\phi'}^2,
\label{calHp}
\end{eqnarray}
where the prime denotes a derivative with respect to $\eta_E$, ${\cal
  H}:= a'/a$ and $\partial V:=dV(\phi)/d\phi$.  We note that the above
three equations are not independent. One of them can be derived from
the other two. The bounce solution is the solution that satisfies the
above set of equations with the boundary condition that $\phi'\to0$,
$a\to0$ and ${\cal H}\to \pm1$ at $\eta_E\to\mp\infty$.  The topology
of the solution is the 4-sphere.  Hereafter, we refer to scalar field
and the scale factor of the background bounce solution by $\phi$ and
$a$.  The two points at which $a=0$ are the centers of the
$O(4)$-symmetry.  For convenience, we take $\eta_E=\infty$ to be on
the true vacuum side (i.e., inside the $O(4)$-symmetric bubble), and
call it the pole and that on the false vacuum side the antipole.

Sometimes it is convenient to use
the coordinate $\tau_E$ defined by $d\tau_E=a(\eta_E)d\eta_E$.
In terms of it, the background equations are written as
\begin{eqnarray}
 && \ddot\phi+3{\dot a\over a}\dot\phi-{\partial V}=0,
\label{phidd}\\
 && \left({\dot a\over a}\right)^2-{1\over a^2}
 ={\kappa\over 3}\left({1\over 2}{\dot\phi}^2-V(\phi)\right), 
\label{Friedmann}\\
 && \left({\dot a\over a}\right)^{\displaystyle\cdot}+{1\over a^2}
 = -{\kappa\over 2}{\dot\phi}^2,
\label{Hdoteq}
\end{eqnarray}
where the dot represents a derivative with respect to $\tau_E$.

\begin{figure}
\centerline{\epsfbox{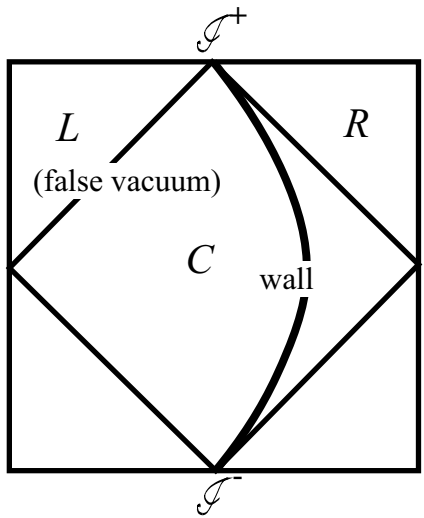}}
\caption{Conformal diagram of a de Sitter-like space with a bubble.}
\label{fig5}
\end{figure}

Once we know the bounce solution, 
the background geometry and the field configuration 
in the Lorentzian region 
are obtained by the analytic continuation of the bounce solution
through the 3-sphere intersecting both poles.
The coordinates in the Lorentzian region are given by
\begin{eqnarray}
 && \eta_E=\eta_C=-\eta_R-{\pi\over 2}i=\eta_L+{\pi\over 2}i\,, \cr
&& \chi_E=-i\chi_C+{\pi\over 2}=-i\chi_L=-i\chi_R\,, \cr
 && a=a_C=i a_R=i a_L\,. 
\label{coordac}
\end{eqnarray}
Accordingly, the analytic continuation of the coordinate $\tau_E$ is
given by 
\begin{equation}
  d\tau_E=d\tau_C=-id\tau_R=id\tau_L\,.
\end{equation}
The coordinates with the indices $C$, $R$ and $L$ cover
regions $C$, $R$ and $L$, respectively, in Fig.~\ref{fig5}.
The region $R$ corresponds to the inside of the lightcone 
emanating from the center of the nucleated bubble, where $a=0$ and
$\eta_C=\eta_E\to +\infty$. 
The region $L$ is the inside of the lightcone emanating from the
antipodal point where $a=0$ and $\eta_C\to -\infty$.
The region $C$ is the remaining central region.
In region $C$, the metric is given by 
\begin{eqnarray}
 ds^2=a(\eta_C)^2\left[d\eta_C^2-
 d\chi_C^2+\cosh^2\chi_C(d\theta^2+\sin^2\theta d\varphi^2)\right].
\label{metricc}
\end{eqnarray}
One sees that surfaces that respect the maximal symmetry, i.e., 
the $\eta_C=$const. hypersurfaces are no longer spacelike.
Instead of $\eta_C$, the coordinate $\chi_C$ plays the role of the time
coordinate there. Note that the $\chi_C=$const. hypersurfaces are 
not homogeneous.

In order to find the quantum state after the bubble nucleation, 
we need to find a set of normalized mode functions that
satisfies the regularity condition explained in Introduction. 
As mentioned there and can be seen from Fig.~\ref{fig5},
no Cauchy surface exists in regions $R$ and $L$. Thus, in order to
obtain a set of normalized mode functions, 
we need to find the reduced action in region $C$, 
starting from the second order variation of 
the action including the metric perturbation on this background. 
This was done in our previous paper\cite{prev}.  
There, we used the Faddev-Jackiw method for constrained systems
to obtain the reduced action. In Appendix A of this paper, as an
alternative and perhaps more familiar method, we present a derivation
of the reduced action a la Dirac.
Perturbations on a spherical symmetric background can be decomposed 
into even and odd parity modes. The odd parity modes do not
contribute to the CMB anisotropy if we choose the position of the
observer as the center of the spherical symmetry.
Hence we concentrate on even parity modes. 
As mentioned above, we must work in region $C$, 
where the background configuration is spatially 
inhomogeneous. However, in region $R$ or $L$, the 
background solution has the symmetry of the FRW universe.
There the standard cosmological 
perturbation theory tells us that even parity perturbations 
for the Einstein-scalar system 
can be decomposed into the scalar and tensor-type 
perturbations.
Therefore, one expects that even 
parity modes can be decomposed into two sets of decoupled 
perturbations even in region $C$ and they are naturally
identified with the scalar and tensor-type perturbations, respectively,
when they are analytically continued to region $R$ or $L$. 
This was shown to be true in \cite{prev}. Hence, following the
conventional terminology used in the cosmological perturbation theory, 
we call the corresponding modes the scalar-type perturbation 
and the tensor-type perturbation, respectively, even in region $C$. 

In this section, we work in region $C$ where the Cauchy surface
exists. Note that we have $\eta_E=\eta_C$ and $\tau_E=\tau_C$.
For notational simplicity, we omit the index $C$ from the coordinate 
variables throughout this section.

\subsection{Reduced action for scalar-type perturbation}
\label{actsca}

First we consider the scalar-type perturbation. 
We introduce a gauge invariant variable $Q_S$ for the scalar-type
perturbation which corresponds to the curvature perturbation in the
Newton gauge in the open universe. Explicitly the metric and the scalar
field perturbation in the Newton gauge is given as
\begin{eqnarray}
 ds^2 & = & \left(1+{Q_S\over 4a}\right) d\tau^2 
   +a^2\left(1-{Q_S \over 4a}\right) 
    \left(-d\chi^2 +\cosh^2\chi d\Omega^2 \right), 
\cr 
 \varphi_N & = & {\dot Q_S\over 4\kappa a\dot\phi}\,. 
\end{eqnarray}
In terms of $Q_S$, the reduced action is obtained in Appendix A as
\begin{equation}
 S^{(2)}_{\hbox{sca}}=\sum_{\ell,m}
   \int d\chi \int {d\tau \over 2a^3 \dot\phi^2} 
   \left[ \cosh^2\chi {\partial\over \partial \chi}
         \overline{Q_S^{\ell m}}
          \widehat K_S{\partial\over \partial \chi}{Q_S^{\ell m}}
   -\overline{Q_S^{\ell m}}\widehat K_S 
      \left\{{\ell(\ell+1)}
+(\widehat K_S -3)\right\}{Q_S^{\ell m}}\right],
\label{Ssca}
\end{equation}
where $\ell$ and $m$ are the eigenvalues of spherical harmonic 
expansion, $Q_S^{\ell m}$ is the $(\ell,m)$-component 
of the spherical harmonic expansion, 
$Q_S=:\sum Q_S^{\ell m} Y_{\ell m}(\Omega)$, 
and $\widehat K_S$ is a derivative operator given by 
\begin{equation}
 \widehat K_S=\left[
  -a^3\dot\phi^2{\partial\over\partial\tau}{1\over a\dot\phi^2}
   {\partial\over\partial\tau}+{\kappa a^2 \dot\phi^2\over 2}\right]. 
\end{equation}
The relation of $Q_S$ to the variable $\bq$ introduced in \cite{prev} is 
\begin{equation}
 Q_S= 4 \kappa a\dot\phi{\bbox q}\,. 
\label{QStobq}
\end{equation}
Putting $\bq=\sum\bq^{\ell m}Y_{\ell m}$, we can use this relation to
rewrite the reduced action (\ref{Ssca}) into the form, 
\begin{equation}
 S^{(2)}_{\hbox{sca}}=\sum_{\ell,m}
      {1\over 2}\int d\chi\int {d\eta} 
     \left[\cosh^2\chi 
       {\partial\overline{{\bbox q}^{\ell m}}\over \partial \chi}
         \widehat{\cal O}{\partial{\bbox q}^{\ell m}\over \partial \chi}
      -\overline{{\bbox q}^{\ell m}}\widehat{\cal O} 
      \left\{{\ell(\ell+1)}+(\widehat{\cal O}-3)
         \cosh^2\chi \right\}{\bbox q}^{\ell m}\right], 
\label{actpre}
\end{equation}
where the operator $\widehat{\cal O}$ is given by
\begin{equation}
 \widehat{\cal O}=-{d^2\over d\eta^2}+{\kappa\over 2}{\phi'}^2 
           +\phi'\left({1\over \phi'}\right)''.
\label{defOhat}
\end{equation}
The reduced action (\ref{actpre}) exactly coincides with the one
derived in \cite{prev}. Notice that in terms of $Q_S$ the action
contains the operator $\widehat K_S$, which is not Schr\"odinger-like,
whereas in terms of $\bbox q$ it contains the operator $\widehat{\cal O}$,
which is. Although this is of no essential significance, the second
form is more convenient when normalizing the modes and when discussing
the general form of the primordial spectrum.

Now let us construct a set of positive frequency functions 
corresponding to the variable $\bq$. 
It is decomposed as
\begin{equation}
{\bq}^{\ell m}=\sum {\cal N}^S_p \bq^{p}(\eta) f^{p\ell}(\chi),
\label{qexpa}
\end{equation} 
where ${\cal N}_p^S$ is a normalization constant, and the
sum is understood as a sum over the discrete spectra and as an
integral over the continuum one. In this expression, $\bq^{p}(\eta)$
is a spatial eigenfunction of the operator $\widehat{\cal O}$ with the
eigenvalue $p^2+4$.  That is, $\bq^p(\eta)$ satisfies
\begin{equation} 
\widehat{\cal O}{\bq}^p
= \left[-{d^2\over d\eta^2}+{\kappa\over 2}{\phi'}^2 
           +\phi'\left({1\over \phi'}\right)''\right]
{\bq}^p=(p^2+4){\bq}^p.  
\label{qeq}
\end{equation}

The equation that determines the time evolution 
is found to be model independent and becomes 
\begin{equation}
 \left[-{1\over\cosh^2\chi}{\partial\over \partial\chi}
    \cosh^2\chi{\partial\over \partial\chi}
    -{\ell(\ell+1)\over \cosh^2\chi}\right] f^{p\ell} 
  =(p^2+1) f^{p\ell}. 
\label{equationf}
\end{equation}
Aside from the normalization factor, the positive frequency function
is determined by the Bunch-Davies regularity condition, namely, it
must be regular at $\chi=0$ ($\chi_E=\pi/2$). The solution is
\begin{equation} 
 f^{p\ell}(\chi)=\sqrt{{\Gamma (ip+\ell +1)\Gamma (-ip+\ell +1)
   \over \Gamma(ip)\Gamma(-ip)\cosh \chi}}
   P^{-\ell-1/2}_{ip-1/2} (i\sinh \chi),
\label{solutionfc} 
\end{equation}
where $P^{\mu}_{\nu}$ is the associated Legendre function 
of the first kind, and we have fixed the normalization factor so that
the analytic continuation of 
$Y^{p\ell m}(\chi,\Omega):=f^{pl}(\chi) Y_{\ell m}(\Omega)$ to region 
$R$ or $L$ satisfies 
\begin{equation}
\int \sinh^2\chi \,d\chi \,d\Omega\, Y^{p\ell m}\overline{Y^{p'\ell'm'}}
  = \delta(p-p')\delta_{\ell\ell'}\delta_{mm'}, 
\label{normY}
\end{equation}
for $p^2>0$, where $\chi=\chi_R$ or $\chi_L$. Also, these 
analytically continued harmonics satisfy
\begin{equation}
{}^{(3)}\Delta\,Y^{p\ell m} = -(p^2+1) Y^{p\ell m},
\label{laplace}
\end{equation}
where ${}^{(3)}\Delta$ is the Laplacian on the unit spacelike
hyperboloid.  For $p^2<0$, the normalization factor in Eq.
(\ref{solutionfc}) is not suitable. In this case, the solutions to
(\ref{equationf}) will be denoted by $f^{\Lambda,\ell}(\chi)$, where
$\Lambda^2=-p^2$. The positive frequency functions
$f^{\Lambda,\ell}(\chi)$ are taken as the usual Bunch-Davies ones,
\begin{equation}
f^{\Lambda,\ell}(\chi) = \sqrt{{\Gamma (\Lambda+\ell +1)
    \Gamma (-\Lambda+\ell +1)
   \over 2 \cosh \chi}}
   P^{-\ell-1/2}_{\Lambda-1/2} (i\sinh \chi),
\end{equation} 
and we define ${\cal Y}^{\Lambda,\ell m}(\chi,\Omega) := f^{\Lambda,\ell}(\chi)
Y_{\ell m}(\Omega)$. Notice that with this choice of positive
frequency functions, ${\cal
  Y}^{\Lambda,\ell m}$ are already Klein-Gordon normalized in the (2+1)
dimensional sense, i.e.
\begin{equation}
i\cosh^2\chi \left\{
 {\partial f^{\Lambda,\ell}\over\partial\chi}\overline{f^{\Lambda,\ell}}
 - f^{\Lambda,\ell} {\partial \overline{f^{\Lambda,\ell}}
   \over\partial\chi}\right\}\int d\Omega\, Y_{\ell m}
 \overline{Y_{\ell'm'}}=\delta_{\ell\ell'} \delta_{mm'}\,.
\end{equation}
These mode functions ${\cal Y}^{\Lambda,\ell m}$ also satisfy
Eq. (\ref{laplace}) in region R.  

In order to quantize the perturbations, it is convenient to write the
action (\ref{Ssca}) in the canonical form (\ref{action}).  Introducing
a new variable $\check\bq:=\sum \sqrt{p^2+4}\, a^{-1} {\cal N}^S_p
\bq^{p}(\eta) Y^{p\ell m}(\chi,\Omega)+\sum \sqrt{4-\Lambda^2}\,a^{-1}
{\cal N}^S_{,\Lambda} \bq^{,\Lambda}(\eta) {\cal Y}^{\Lambda,\ell
  m}(\chi,\Omega)$, the action for $\check\bq$ turns out to be the one
for a scalar field with $\eta$-dependent mass.  Thus the normalized
positive frequency functions are determined by calculating the
ordinary Klein-Gordon norm for $\check\bq$. Namely we require
\begin{eqnarray}
&& i({p^2+4})\cosh^2\chi \vert{\cal N}_p^S \vert^2 \left\{
 {\partial f^{pl}\over\partial\chi}\overline{f^{pl}}
 - f^{pl} {\partial \overline{f^{pl}}\over\partial\chi}\right\}
\int_{-\infty}^{\infty} d\eta\, \bq^{p}\overline{\bq^{p'}} 
\cr &&\quad = {2p(p^2+4) \sinh\pi p\,\vert{\cal N}_p^S \vert^2\over \pi} 
\int_{-\infty}^{\infty} d\eta\, \bq^{p}\overline{\bq^{p'}} 
=\delta(p-p')
%\delta_{p,p'},  
\label{normalS}
\end{eqnarray}
for the continuous spectrum ($p^2>0$), and
\begin{eqnarray}
\label{normsupB}
({4-\Lambda^2})\vert{\cal N}_{,\Lambda}^S\vert^2
\int_{-\infty}^{\infty} 
d\eta\, \bq^{,\Lambda}\overline{\bq^{,\Lambda'}}=\delta_{\Lambda\Lambda'},
\end{eqnarray}
for the discrete one ($p^2<0$).
%where $\delta_{p,p'}$ is either the Dirac Delta function 
%for continuous spectrum ($p^2>0$) or the Kronecker's delta for discrete
%spectrum ($p^2<0$).  
Notice that the Klein-Gordon normalization reduces to a
Schr\"odinger-like normalization for ${\bbox q}_p$ (once a convenient
choice of ${\cal N}_p^S$ has been made) 

The variable $Q_S$ can be also decomposed in the same way as $\bq$. 
{}From Eq.~(\ref{QStobq}), the corresponding spatial eigenfunction
$Q_S^p(\tau)$ is related to $\bq^p(\tau)$ as
$Q_S^p=\sqrt{\kappa}\,a\dot\phi\,\bq^p$. The equation satisfied by 
$Q_S^p$ is given by the eigenvalue equation for the operator 
$\widehat K_S$ as
\begin{equation}
a^{-2}\widehat K_S Q_S^p=
 \left(-a\dot\phi^2{d\over d \tau}{1\over a\dot\phi^2}{d\over d\tau}
     +{\kappa\over 2}\dot\phi^2\right) {Q_S^{p}}={p^2+4\over a^2}Q_S^p. 
\label{Peq}
\end{equation}

For later convenience, we also write down the equation for the
scalar field fluctuation in the Newton gauge:
\begin{eqnarray}
 \ddot\varphi_N^p+3{\dot a\over a}\dot\varphi_N^p+
 \left({p^2+1\over a^2}-\partial^2V\right) \varphi_N^p
 =2\kappa\dot\phi^2\varphi_N^p
+{d\over d\tau}(a\dot\phi)\,{Q_S^p\over4a^2}\,,
\label{phieq}
\end{eqnarray}
where $\partial^2V:=d^2V/d\phi^2$ and $\varphi_N$ has been expanded in the
same way as ${\bbox q}$. 
We note that from the Newton gauge condition~(\ref{Gcondi}) in Appendix A,
we have
\begin{equation}
   \dot Q_S^p=4\kappa a\dot\phi \varphi_N^p
= 4 \kappa{d\over d\tau}(a\dot\phi\,\bq^p).
\label{dotP}
\end{equation}
Using this, Eq.~(\ref{phieq}) may be rewritten as
\begin{equation}
 \ddot\varphi_N^p+3{\dot a\over a}\dot\varphi_N^p
 \left({p^2+1\over a^2}-\partial^2V\right) \varphi_N^p
 =3\kappa\dot\phi^2\varphi_N^p
-\kappa \dot\phi^2\dot\bq^p.
  \label{phieqnew}
\end{equation}

The analytic continuation of the variables $Q_S$, $\varphi_N$
and $\bq$ to region $R$ or $L$ is given by
\begin{equation}
 Q_S^C=i Q_S^R=i Q_S^L,  \quad\varphi_N^C=\varphi_N^R=\varphi_N^L,
\quad {\bq}_C=-i {\bq}_R=i\bq_L, 
\end{equation}
where the indices $C$, $L$ and $R$ denote the variables in the
respective regions.  The evolution equation in an open universe inside
the bubble is given by the analytic continuation of Eq.~(\ref{qeq}),
or Eq.~(\ref{phieqnew}) supplemented by Eq.~(\ref{dotP}).  We note
that the form of the equations does not change after the analytic
continuation except for the signature in front of $\partial^2V$ in
Eq.~(\ref{phieqnew}).

\subsection{Reduced action for tensor-type perturbation}
\label{actten}

In this subsection, we consider the tensor-type perturbation.  We
introduce a gauge invariant variable $Q_T$ that corresponds to the
tensor-type perturbation in the transverse-traceless gauge in the open
universe. Let us denote the metric on a unit 2-sphere by
$d\Omega^2=\hat\sigma_{AB}d\sigma^A d\sigma^B$, and the covariant
derivative and the Laplacian with respect to $\hat\sigma_{AB}$ by
${}^{(2)}\nabla$ and ${}^{(2)}\Delta$, respectively.  The tensor-type
perturbation is expressed as
\begin{eqnarray}
 ds^2 & = & d\tau^2-\left(a^2-{2Q_T\over \cosh^2\chi}\right)d\chi^2
    +\left(a^2+{Q_T\over \cosh^2\chi}\right)\cosh^2\chi\,d\Omega^2
\cr &&
 +4\left(\partial_\chi+\tanh\chi\right)
 {}^{(2)}\nabla_A({}^{(2)}\Delta)^{-1} Q_T\,d\chi\, d\sigma^A
\cr
&& +4\cosh^2\chi
 \left(2-\widehat K_T+{({}^{(2)}\Delta)-2\over 2\cosh^2\chi}+
       \tanh\chi\partial_{\chi}\right)
\cr
&&\quad\times \left[{}^{(2)}\nabla_A {}^{(2)}\nabla_B 
      -{1\over 2}\hat\sigma_{AB}{}^{(2)}\Delta\right]
  ({}^{(2)}\Delta({}^{(2)}\Delta+2))^{-1} Q_T d\sigma^A d\sigma^B, 
\cr
 & =: &ds^2_{(0)} + a^2\,\bh_{ij}[Q_T] dx^{i} dx^{j}, 
\end{eqnarray}
where $i,\, j=\chi,\,\theta,\,\varphi$ and
\begin{equation}
\widehat K_T = \left[-a{\partial\over\partial\tau}
   a^3{\partial\over\partial\tau}{1\over a^2}\right]. 
\end{equation}
The action for $Q_T$ is obtained in Appendix A 
and it is given by 
\begin{eqnarray}
 S^{(2)}_{\hbox{\small GW}}=\sum_{\ell,m}
    {2\over \kappa(\ell-1)\ell(\ell+1)(\ell+2)}
        \int d\chi \int {d\tau \over 2a^3} 
   &&\Biggl[\cosh^2\chi {\partial\over \partial \chi}
               \overline{Q_T^{\ell m}}
          {\widehat K_T(\widehat K_T-1)}
        {\partial\over \partial \chi}{Q_T^{\ell m}}
\cr &&
    -\overline{Q_T^{\ell m}}\widehat K_T (\widehat K_T-1)
      \left\{{\ell(\ell+1)}+\widehat K_T \cosh^2\chi\right\}
      {Q_T^{\ell m}}\Biggr],
\label{SGW}
\end{eqnarray}
where $Q_T^{\ell m}$ is defined 
by $Q_T =:\sum Q_T^{\ell m}Y_{\ell m}$, as before. 

Again, the operator $\widehat K_T$ is not of Schrodinger type.
Therefore, it is convenient to change to the variable $\bw$ introduced
in \cite{prev}, which is related to $Q_T$ by
\begin{equation}
 \bbox{w}={a^2\over \kappa}{\partial\over \partial \tau}{Q_T\over a^2}.
\label{bwtoQT}
\end{equation}
Using this relation, the reduced action (\ref{SGW}) can be rewritten 
as 
\begin{eqnarray}
\label{eq:mainoutvec}
  S_{\hbox{\small GW}}^{(2)} =\sum_{\ell,m}
      {\kappa\over (\ell-1)\ell(\ell+1)(\ell+2)}
     \int d\chi\int {d\eta} 
     &&\Biggl[\cosh^2\chi 
       {\partial\overline{\bbox{w}^{\ell m}}\over \partial \chi}
         \widehat{K}{\partial \bbox{w}^{\ell m}\over \partial \chi}
\cr &&
      -\overline{\bbox{w}^{\ell m}}\widehat{K} 
      \left\{{\ell(\ell+1)}+(\widehat K+1)
         \cosh^2\chi \right\}\bbox{w}^{\ell m}\Biggr], 
\label{actpreGW}
\end{eqnarray}
where the operator ${\widehat K}$ is given by
\begin{equation}
  {\widehat K}= -\frac{\partial^2}{\partial\eta^2} 
       +\frac{\kappa}{2} {\phi'}^2.
\label{Optensor}
\end{equation}
This coincides with the reduced action obtained in \cite{prev}.

Now we construct a set of positive frequency functions. 
First we use the variable $\bw$. As before, we decompose $\bw$ as
\begin{equation}
\bw=\sum \bw^{\ell m}Y_{\ell m}\,,\quad
{\bw}^{\ell m}=\sum {\cal N}^T_{p\ell} \bw^{p}(\eta) f^{p\ell}(\chi). 
\end{equation} 
Note that the time-dependent part, $f^{p\ell}(\chi)$, 
is the same as in the scalar-type perturbation. 
In this expression, the spatial function 
$\bw^{p}(\eta)$ satisfies the eigenvalue equation for the operator
$\widehat{K}$ with the eigenvalue $p^2$:
\begin{equation} 
{\widehat K}\bw^p
= \left[-{d^2\over d\eta^2}+{\kappa\over 2}{\phi'}^2 \right]
{\bw}^p=p^2{\bw}^p.
\label{weq}
\end{equation}
It is important to note that the spectrum of $p$ is continuous
with $p^2>0$ because the potential term ${\kappa\over2}{\phi'}^2$ is
positive definite and vanishes at infinity. Then, defining a new variable 
$\check\bw:=\sum \sqrt{2 \kappa p^2\over 
(\ell-1)\ell(\ell+1)(\ell+2)}\, a^{-1} 
{\cal N}_{p\ell}^T\bw^{p}(\eta) Y^{p\ell m}(\chi,\Omega)$, 
the action reduces to the one for a scalar field, as in the case of
scalar-type perturbation. Therefore, the normalization condition becomes 
\begin{equation}
 {2\kappa p^2\over (\ell-1)\ell(\ell+1)(\ell+2)} 
 {2p\sinh\pi p\, \vert {\cal N}_{p\ell}^T \vert^2\over \pi} 
\int_{-\infty}^{\infty} d\eta\, \bw^{p}\overline{\bw^{p'}} 
=\delta(p-p'). 
\label{normalT}
\end{equation}

Since what we are interested in is the metric perturbation, we now
consider $Q_T$. It can be also decomposed as
 ${Q_T}=\sum {\cal N}_{p\ell}^T {Q_T}^p(\tau)Y^{p\ell m}(\chi,\Omega)$, 
where the spatial function $Q_T^p$ satisfies the eigenvalue equation
for the operator $\widehat K_T$:
\begin{equation}
\widehat K_T{Q_T^p}=
- a^4\left[{d^2\over d\tau^2}+3{\dot a\over a}{d\over d\tau} \right]
{Q_T^p\over a^2}= (p^2+1)Q_T^p\,.
\label{QTeq}
\end{equation}
Then the metric perturbation $\bh_{ij}[Q_T]$ is given by
\begin{equation}
  \label{Tmetpert}
  \bh_{ij}[Q_T]= 2
\sum\sqrt{2p^2(p^2+1)\over (\ell-1)\ell (\ell+1)(\ell+2)}\,
{\cal N}_{p\ell}^T \frac{Q_T^p}{a^2} Y_{ij}^{(+)p\ell m},
\end{equation}
where 
\begin{equation}
 Y_{ij}^{(+)p\ell m}:= {1\over 2}
 \sqrt{(\ell-1)\ell (\ell+1)(\ell+2)\over 2p^2(p^2+1)} 
  \,a^2\,\bh_{ij}[Y^{p\ell m}].
\label{defYij}
\end{equation}
The tensor $Y_{ij}^{(+)p\ell m}$ is defined so that its 
analytic continuation to region $R$ or $L$ gives the normalized even
parity tensor harmonics on the unit 3-hyperboloid \cite{tomita}, i.e.,
\begin{equation}
 \int d^3x \sqrt{\hat\gamma}\, \hat\gamma^{ii'}\hat\gamma^{jj'}
  Y^{(+)p\ell m}_{ij} \overline{Y^{p'\ell' m'}_{i'j'}}
  = \delta(p-p') \delta_{\ell\ell'}\delta_{mm'}\,,
\end{equation}
where $\hat\gamma_{ij}$ is the metric on the unit 3-hyperboloid
and $\hat\gamma=\det \hat\gamma_{ij}$. 
Therefore it is more relevant to introduce a new normalization constant 
for the tensor perturbation:
\begin{equation}
  \tilde{\cal N}_{p}^T:
= 2 \sqrt{2p^2(p^2+1)\over (\ell-1)\ell (\ell+1)(\ell+2)}\,
{\cal N}_{p\ell}^T\,,
\label{newnormT}
\end{equation}
with which the metric perturbation is expressed as
\begin{equation}
\bh_{ij}[Q_T]= 
\sum\tilde{\cal N}_{p}^T \frac{Q_T^p}{a^2} Y_{ij}^{(+)p\ell m}.
\label{defbh}
\end{equation}
In terms of $\tilde{\cal N}_p^T$, the normalization condition
(\ref{normalT}) is re-expressed as
\begin{equation}
  {\kappa p\sinh\pi p  \vert\tilde{\cal N}_{p}^T\vert^2 
       \over 2 \pi(p^2+1)}
   \left[\int_{-\infty}^{\infty} d\eta \bw^{p} 
   \overline{\bw^{p'}}\right]=\delta(p-p').
\label{normtilT}
\end{equation}
The relations between the analytically continued variables 
are given by 
\begin{equation}
 Q_T^C=Q_T^R=Q_T^L, \quad 
 \bw_C=i\bw_R=-i\bw_L.
\end{equation}

\section{scalar supercurvature modes}
\label{secsuper}
The possible presence of scalar supercurvature modes (i.e. modes with
$p^2<0$) has important physical consequences.  If a model has
supercurvature scalar modes then the density parameter $\Omega$
becomes a random variable which takes different values in different
places of the universe, and as we shall see, it may have a
considerable spread if $p^2$ is close to $-1$ \cite{modelq,anth}.
Also, if the spectrum of perturbations around a particular bubble
contains a ``supercritical'' eigenvalue with $p^2<-1$, this may
indicate that we are not looking at the dominant channel for decay,
and that there is another instanton with a lower Euclidean action
(This is known as the ``no supercritical supercurvature'' mode
conjecture which has been shown to hold in several cases
\cite{supcrit}). In this section we shall analyze the conditions for
the existence of scalar supercurvature modes in one field models of
open inflation. We conclude that generically, these will be absent in
the thin wall case, although they can be present for thick wall models
(and even for thin wall models with sufficiently complicated
potentials.) Intuitively, the reason is that for thin wall models the
microphysical mass scales are large compared with the inverse size of
the bubble, and hence all modes of fluctuation are ``hard''. In
contrast, two-field models generically have supercurvature modes
\cite{YST} even in the thin wall case. These are not due to the
tunneling field but to the ``soft'' modes of the slow roll inflaton
field - roughly speaking, the value of the slow roll field at the time
of nucleation may be seen as an approximate zero mode of the action.

First of all, in subsection \ref{subY} we shall investigate the weak
gravity limit, comparing the scalar spectrum in the case when metric
perturbations are included and the case when they are neglected.  The
formalism of supersymmetric quantum mechanics turns out to be very
useful in this respect. Then, in subsection \ref{subZ} we shall give a
sufficient condition for the existence of scalar supercurvature modes.
In subsection \ref{subA}, we describe a method to construct a model
which has supercurvature modes when the bubble wall is thick but does
not have them when the bubble wall is thin. In subsection \ref{subBB}
we consider the generalization of the Fubini instanton to a de Sitter
background and use it to illustrate the role played by supercurvature
modes in one field models.

\subsection{Supersymmetric Quantum Mechanics and the weak gravity limit}
\label{subY}

For scalar perturbations the spectrum is determined by the Schr{\"o}dinger 
operator $\widehat{\cal O}$ in Eq. (\ref{qeq}). This can be rewritten
in the form  
$$
\widehat{\cal O} {\bbox q}^p 
= [\widehat{\cal O}_0 + (\kappa \phi'{}^2/2)]{\bbox q}^p = 
(p^2 + 4) {\bbox q}^p, 
$$ 
where
\begin{equation}
\widehat{\cal O}_0 
= - {\partial^2 \over \partial \eta^2} - W' + W^2,
\label{osusy}
\end{equation}
and we have defined
$$W= \phi''/\phi'.$$ 
{}From the boundary conditions satisfied by the instanton, it is easy to 
show that
\begin{equation}
\phi' \to e^{\pm 2 \eta}\quad (\eta \to \mp \infty)
\label{boundary0}
\end{equation}
Hence, we have 
\begin{equation}
W\to \pm 2 \quad (\eta \to \mp \infty),
\label{boundary}
\end{equation} 
and the potential term in (\ref{osusy}) tends to 4 at infinity.
Therefore the spectrum is continuous for $p^2>0$. Also, since the
potential is not bounded below by its asymptotic value, there may be a
discrete spectrum of supercurvature modes for $p^2<0$. The
corresponding harmonics ${\cal Y}^{\Lambda,lm}$ have a wavelength
longer than the curvature scale in the open universe (and are not
normalizable on the $t=\text{const.}$ hyperboloids \cite{opendeS}.)

Let us now consider the spectrum of perturbations of the scalar field
$\Phi$ when the metric perturbations are switched off. This is the
limit where kinetic terms of the form $\kappa \phi'{}^2$ are neglected,
while the effective cosmological constant $\kappa V$ is kept finite.
In that case, the background geometry is de Sitter and the equation
for scalar field perturbations is $ \Bigl[\nabla^\mu\nabla_\mu
-\partial^2V(\eta_C)\Bigr]\varphi_ {\bf k}(\eta,{\bf x})=0.  $ Using
the coordinates (\ref{metricc}) and substituting the ansatz
\begin{equation}
\varphi=\sum  {{\bbox \chi}^p(\eta) \over a} Y^{p\ell m}({\bf x}),
\label{eim}
\end{equation}
where $Y^{p\ell m}$ is understood as ${\cal Y}^{\Lambda,\ell m}$ for
$p^2<0$, one readily finds the eigenvalue equation for ${\bbox
  \chi}^p$:
\begin{equation}
 -{\partial^2 \over \partial \eta^2} {\bbox \chi}^p +
           [a^2 \partial^2 V + {\cal H}'+ {\cal H}^2 + 3] {\bbox \chi}^p 
          =(p^2+4) {\bbox \chi}^p.
\label{ewq}
\end{equation}
Using the scale factor for de Sitter space $a(\eta)=1/(H\cosh\eta)$, 
this can be rewritten in the form
\begin{equation}
 -{\partial^2 \over \partial \eta^2} {\bbox \chi}^p +
                   [a^2 (\partial^2 V-2 H^2) + 4] {\bbox \chi}^p = 
                   (p^2+4) {\bbox \chi}^p.
\label{ewq2}
\end{equation}
Note that in de Sitter space the background field equation becomes 
\begin{equation}
\phi''-2\tanh\eta\,\phi' -{1\over H^2\cosh^2\eta}\partial V(\phi)=0.
\label{phiNBR}
\end{equation}
Taking the $\eta-$derivative after multiplying it by $\cosh^2\eta$, we
have 
\begin{eqnarray}
 \left[{d^2\over d\eta^2}-
\left(4+{\partial^2 V -2 H^2\over H^2 \cosh^2\eta}\right)
 \right] \phi'=0.
\label{prime1}
\end{eqnarray}
It is clear from Eq. (\ref{prime1}) that 
Eq. (\ref{ewq2}) has as its lowest eigenfunction
the wall fluctuation mode ${\bbox \chi}^{p=2i} = \phi'$, with eigenvalue
$p^2=-4$. Hence, we can rewrite the operator in the left hand side
of (\ref{ewq2}) as
\begin{equation}
\widehat{\cal P} 
= -{\partial^2 \over \partial \eta^2} + {\phi'''\over \phi'} 
         = -{\partial^2 \over \partial \eta^2} + W' + W^2,
\label{pop}
\end{equation}
where $W$ is defined as in (\ref{osusy}). In going from (\ref{ewq})
to (\ref{pop}) we have used that the background metric is de Sitter.
However, for future reference, we take (\ref{pop}) as the 
definition of the operator $\widehat{\cal P}$, even in the case of
strong gravity. 

Thanks to the formalism of 
supersymmetric quantum mechanics \cite{susy},
the correspondence between the spectrum of 
$\widehat{\cal P}$ and that of $\widehat{\cal O}$ in the limit of 
weak gravity becomes rather straightforward.
It turns out that $\widehat{\cal O}_0$ and $\widehat{\cal P}$ are
supersymmetric partners 
of each other, so neglecting the term $\kappa\phi'{}^2$, both 
spectra coincide (except for the ground state). Indeed, in terms of the 
``supersymmetry'' generators 
$$\widehat Q=\partial_{\eta}+W, \quad 
\widehat Q^{\dagger}=-\partial_{\eta}+W,$$ 
we have 
$$\widehat{\cal O}_0=\widehat Q^{\dagger}\widehat Q, \quad
\widehat{\cal P}=\widehat Q\widehat Q^{\dagger}.$$ 
Note that if supercurvature modes are absent for the operator
$\widehat{\cal O}_0$, they are absent for $\widehat{\cal O}$ since
the neglected term $\kappa\phi'{}^2/2$ is positive definite.
Normalized eigenstates ${\bbox \chi}^p_S$ of $\widehat{\cal P}$
 (which we may call ``bosons'') 
can be transformed into normalized eigenstates ${\bbox q}_S^p$ of
$\widehat{\cal O}_0$  
(``fermions'') and vice versa by acting upon them with the SUSY generators:
\footnote{The subindices $S$ are included to indicate that these are 
eigenstates of the SUSY related operators $\widehat{\cal O}_0$ and
$\widehat{\cal P}$, and 
not of physical operators such as $\widehat{\cal O}$.}
\begin{equation}
\widehat Q^{\dagger} {\bbox \chi}^p_S = (p^2+4)^{1/2} {\bbox q}^p_S,
\label{susytrans1}
\end{equation}
\begin{equation}
\widehat Q {\bbox q}^p_S = (p^2+4)^{1/2} {\bbox \chi}^p_S.
\label{susytrans2}
\end{equation}
Note that, for convenience, in the above equations we have taken the 
eigenstates to be normalized as in the corresponding Schr{\"o}dinger problem 
[and not as in (\ref{normalS})], i.e.
\begin{eqnarray}
\int_{-\infty}^{\infty}{\bbox q}_S^p \overline{{\bbox q}_S^{p'}}=
\delta_{p,p'},
\hskip 1cm
\int_{-\infty}^{\infty}{\bbox \chi}_S^p \overline{{\bbox
    \chi}_S^{p'}}=
\delta_{p,p'}
\end{eqnarray}

Therefore, the spectra coincide {\em except} for the 
ground state of $\widehat{\cal P}$ with $p^2+4=0$, 
whose eigenfunction ${\bbox \chi}^{p=2i}_S$ is annihilated by 
$\widehat Q^{\dagger}$ and has no fermionic partner. The reason is
that this partner would satisfy the equation $\widehat Q {\bbox q}_S=0$,
which due to the boundary condition 
(\ref{boundary}) has no normalizable solutions. This is why the
wall fluctuation mode disappears from the scalar spectrum when gravity,
even if weak, is turned on. 

Where does the wall fluctuation mode go? Even in the absence of
gravity, we may choose to eliminate the mode ${\bbox \chi}^{p=2i}$
from the scalar spectrum.  Through a coordinate transformation, it can
be rewritten as an even parity tensor mode with $p^2=0$
\cite{TStensor}, which is precisely the bottom of the continuum
spectrum of the operator $\hat K$ corresponding to tensor modes,
defined in Eq.~(\ref{weq}).  When the gravitational coupling is
switched on, it turns out that the effect of wall fluctuations, which
was concentrated in the $p^2=0$ tensor mode, spreads to the long
wavelength region of the spectrum, with eigenvalues \cite{STY}
$$
p^2 \lesssim \int_{-\infty}^{\infty} \kappa \phi'{}^2 d\eta \ll 1.
$$
It has been shown that these long wavelength tensor modes
reproduce the effect of the wall fluctuation mode on the
cosmic microwave background in the limit of small $\kappa \phi'{}^2$ 
\cite{STY}.

In \cite{YST} it was shown that, neglecting metric fluctuations and
under rather generic conditions for the bubble profile in the thin
wall limit, the operator $\widehat{\cal P}$ has the wall fluctuation
mode as its only supercurvature mode. Specifically, the discussion
given there assumes a step function model of $\partial^2V$ that has
the following behavior of $\partial^2V$. First, $\partial^2V-2H^2>0$
on the false vacuum side followed by a thin region of
$\partial^2V-2H^2<0$. Then there is a thin region of
$\partial^2V-2H^2>0$ followed by a region of $\partial^2V\ll 2H^2$ on
the true vacuum side.\footnote{It can be easily seen that letting the
  region of $\partial^2V-2H^2>0$ thick or letting $\partial^2V$ on the
  true vacuum side larger will only make the existence of a
  supercurvature mode more difficult.}  It follows from our previous
discussion that under those conditions the operator $\widehat{\cal O}$
will not have any supercurvature modes.

We can also argue another case for the non-existence of a
supercurvature mode as follows.  Let us consider Eq.~(\ref{ewq2}) for
${\bbox \chi}^p$.  We assume the case when the background solution is
such that a region of $U_{eff}(\eta):=(\partial^2V-2H^2)a^2<0$ exists
only for a finite interval of $\eta$.  Let $\eta_i$ ($i=1,2$) be the
two zero points of $U_{eff}$; $U_{eff}(\eta_1)=U_{eff}(\eta_2)=0$. We
assume $\eta_1<\eta_2$.  Now consider a one-node solution of ${\bbox
  \chi}^p$ in the interval $\eta_1<\eta<\eta_2$ with the boundary
conditions ${\bbox \chi}^p{}'(\eta_1)={\bbox \chi}^p{}'(\eta_2)=0$.
Let $p_*^2$ be the eigenvalue of this solution.  Since $U_{eff}$ is
positive elsewhere, if there exists a one-node solution ${\bbox
  \chi}^{p_0}$ with $p^2_0<0$ for the original problem, ${\bbox
  \chi}^{p_0}$ must be more warped than ${\bbox \chi}^{p_*}$ in the
interval $\eta_1<\eta<\eta_2$. Then it follows that $p_0^2>p_*^2$. In
other words, there is no supercurvature mode if $p_*^2>0$.

As we shall see in subsection~\ref{subA}, there exists a method to
reconstruct a potential $V$ from a given $\phi'$ as a function of
$\eta$ provided it satisfies the boundary condition (\ref{boundary0})
and it is positive (or negative) definite.  Let
$\tilde\phi'(\eta)=\phi'(\alpha\eta)$ where $\alpha$ is a constant
within the interval $\eta_1<\eta<\eta_2$ and $\alpha\to1$ for
$\eta\to\pm\infty$. Then one can always construct a potential $\tilde
V$ from $\tilde\phi'$.  Now let us consider a constant scale
transformation $\eta\to\alpha\eta$ to the equation for ${\bbox
  \chi}^p$ within the interval $\eta_1<\eta<\eta_2$:
\begin{eqnarray}
  \label{chiscale}
(p^2+4){{\bbox \chi}}^p(\alpha\eta)&=&
\left[-{\partial^2\over\partial(\alpha\eta)^2}
+{1\over\phi'(\alpha\eta)}
{\partial^2\over\partial(\alpha\eta)^2}{\phi'(\alpha\eta)}
\right]{{\bbox \chi}}^p(\alpha\eta)
\nonumber\\
&=&{1\over\alpha^2}\left[-{\partial^2\over\partial\eta^2}
+{1\over\tilde\phi'(\eta)}
{\partial^2\over\partial\eta^2}{\tilde\phi'(\eta)}
\right]\tilde{{\bbox \chi}}^{\tilde p}(\eta)
\nonumber\\
&=&{\tilde p^2+4\over\alpha^2}\tilde{{\bbox \chi}}^{\tilde p}(\eta),
\end{eqnarray}
where $\tilde{{\bbox \chi}}^{\tilde p}(\eta)={{\bbox \chi}}^p(\alpha\eta)$.
Thus we see $\tilde{{\bbox \chi}}^{\tilde p}$ is a solution if 
${{\bbox \chi}}^p$ is so, and the eigenvalue $\tilde p$ is given by
\begin{eqnarray}
  \label{tildep}
  \tilde p^2+4=\alpha^2(p^2+4).
\end{eqnarray}
Then, taking ${{\bbox \chi}}^p$ to be the one-node solution with the
eigenvalue $p_*^2$ constructed in the above, the corresponding
one-node solution for the potential $\tilde V$ has the eigenvalue
$$
\tilde p_*^2=\alpha^2(p_*^2+4)-4.
$$
For a sufficiently large $\alpha$, we have $\tilde p_*^2>0$ and
hence there will be no supercurvature mode for $\tilde V$.
Since $\alpha\gg1$ corresponds to taking the thin wall limit, this
implies the supercurvature mode disappears in the thin wall limit
for the class of potentials which are related by the above scale
transformation.

Unfortunately, neither the arguments given in \cite{YST}, nor the above
scaling arguments are completely 
general, and we have not been able to give a proof that thin wall always
implies that there are no supercurvature modes. 
Therefore, in principle,
one has to analyze each case separately. Nevertheless, the results given
in the following subsections may provide some valuable guidance in this
respect.

\subsection{A sufficient condition for the existence of supercurvature
modes and an SWKB estimate}
\label{subZ}

An upper bound on the lowest eigenvalue of the scalar operator 
$\widehat{\cal O}$ can be obtained by the ``variational'' method. 
Writing
$$
\widehat{\cal O} = \widehat{\cal P} - 2 W' + {\kappa \phi'{}^2 \over 2},
$$
we can readily evaluate the expectation value of $\widehat{\cal O}$ in
the ``test'' quantum state whose wavefunction is given by $\phi'$
(i.e. the true ground state of the operator $\widehat{\cal P}$). This
expectation value must be larger than or equal to the lowest eigenvalue
$p_0$ of $\widehat{\cal O}$, which implies 
$$
(p_0^2 +4) \int \phi'{}^2 d\eta \leq  \int \left(4 \phi''{}^2 + 
{\kappa \phi'{}^4 \over 2} \right) d\eta.
$$
Therefore, a sufficient condition for the existence of a supercurvature
mode ($p_0^2<0$) is that the bubble profile satisfies
\begin{equation}
\int \left(\phi''{}^2 - \phi'{}^2 + {\kappa \phi'{}^4 \over 8} \right) d\eta < 0.
\label{suff}
\end{equation}
Typically, the third term in the integrand can be neglected because it is
Planck scale suppressed. Since the first term carries two more derivatives 
than the second, this indicates that we may expect supercurvature modes when 
the bubble walls are thick.

Aside from this sufficient condition, we can also estimate the number of
bound states of $\cal O$ using the supersymmetric version of the WKB 
quantization condition (SWKB) \cite{susy}:
\begin{equation}
\int^a_b \sqrt{p^2+4-W^2} d\eta = \pi n,
\label{SWKB}
\end{equation}
where $n=0,1,2,...$ for the spectrum of the operator $\widehat{\cal P}$, 
and $n=1,2,...$ for the spectrum of the operator $\widehat{\cal O}_0$.
The integration runs over the range where $W^2<p^2+4$.\footnote{Note
  that for the ground state $n=0$ we know that $p^2+4=0$. In this case
  the ``turning points'' are coincident and the SWKB approximation gives
  the exact answer. As noted in \cite{susy}, this approximation gives
  accurate answers at both ends of the allowed range of $n$ ($n=0$ and
  $n\to \infty$), so one may expect better results than in the usual WKB
  approach.} 
Hence, the approximate condition for the existence of a  
supercurvature mode of $\widehat{\cal O}_0$ is
that
\begin{equation} 
I\equiv \int^a_b \sqrt{4-W^2} d\eta > \pi
\label{swkbcond}
\end{equation}
As shown in Appendix B, this condition is not satisfied when the bubble walls 
are thin, so we do not expect supercurvature modes in this case. 

Unlike the sufficient condition (\ref{suff}), however,
the condition (\ref{swkbcond}) is the result of an approximation, and so 
it cannot be used as a rigorous criterion for the existence of 
supercurvature modes.

\subsection{A simple method to construct a model}
\label{subA}

In this subsection we point out that for any given profile $\phi(\eta)$
which is monotonous and has appropriate boundary conditions at infinity,
we can always ``reconstruct'' a potential $V(\phi)$ for which $\phi$ is
a bounce solution. In particular, this freedom can be used to construct 
models which do have supercurvature modes.

Let us consider that $\phi'$ is given. It can be chosen arbitrarily, provided 
that it has no nodes and satisfies (\ref{boundary0}) at infinity. Then, 
the solution of (\ref{calHp}) at $\eta \to \pm\infty$ given by 
$$
 {{\cal H}-1\over {\cal H}+1}=C_{\pm}e^{2\eta}.
 $$
 This solution, which has been obtained from taking $\phi'$
 strictly to zero, automatically satisfies the required boundary
 condition, ${\cal H}\to\mp 1$ as $\eta\to\pm\infty$.  Once
 we know ${\cal H}$, we can determine the scale factor by integrating
 ${\cal H}$. Here an integration constant appears which will determine
 the overall scale of the potential (say the potential at the false
 vacuum $V_F$). By using Eq. (\ref{Friedmann}), the potential is
 determined as a function of $\eta$
\begin{equation}
 V={6-6{\cal H}^2-{\kappa}{\phi'}^2\over 2\kappa a^2}.
\end{equation}
Since by assumption $\phi(\eta)$ is monotonic, we can express $\eta$
in terms of $\phi$ and the shape of the potential $V(\phi)$ is uniquely 
determined up to the overall factor mentioned above. 
\footnote{Note that
the equations of motion for the background have the following symmetry:
a global factor in the potential can be reabsorbed into the scale factor, 
and so its only effect is to change all physical distances by a constant 
factor.}

Here, we note that the positivity of the potential determined in 
this way is not guaranteed in principle. However, integrating the 
equation 
\begin{equation}
 {d\over d\eta}\left({1\over 2}{\dot\phi}^2-V\right)
   =-3{\cal H}{\dot\phi}^2 
\end{equation}
and using the boundary condition $\dot\phi\to 0$ as
$\eta\to\pm\infty$ , we have $\Delta V=3 \int_{-\infty}^{\infty}
a^{-2}{\cal H}{\phi'}^2 d\eta$, where $\Delta V$ is the difference
between the heights of potential in the false vacuum ($\eta\to
-\infty$) and at the nucleation point ($\eta\to \infty$).  If
$\kappa{\phi'}^2$ is sufficiently small, then $\Delta V/V_F$ becomes
small, and hence $V$ stays positive definite.

Having shown that $V$ can be constructed for given $\phi'$, let us now
turn to the eigenvalue equation for the scalar-type 
perturbations in terms of $\phi$, Eq. (\ref{qeq}). If there is an 
eigenfunction with negative $p^2$, 
the solution with $p^2=0$ must have, at least, one node.  
Hence, it is sufficient to examine the equation for $q:=\bq^{p^2=0}$, 
which is written as 
\begin{equation}
\left[-{d^2\over d\eta^2}+{\kappa\over 2} {\phi'}^2+ U(\eta)\right]
q=0, 
\end{equation}
where 
\begin{equation}
 U(\eta)=-W'+W^2-4. 
\end{equation}
Since the change in the overall amplitude of $\phi'$ does not 
alter the function $W$, we can construct a model 
in which the second term is negligibly small without changing the 
potential $U$. 
Hence we neglect this positive definite term. 
As mentioned above, the positivity of $V$ is 
guaranteed in this limiting case. 

Instead of specifying the model by giving $\phi$, we can specify it
by giving $W$, which can be chosen arbitrarily as long as the boundary 
condition (\ref{boundary}) is satisfied. Then, by choosing $W$ to stay near 
zero for a sufficiently long 
time interval, $U$ will have a wide negative-valued region 
and the solution for $q=\bq^{p=0}$ will have a node. 
On the other hand, if the wall is sufficiently thin, $W'$ will 
dominate the potential $U$ when $|W|$ is small. 
Then $U$ may become positive definite, and in this case $q$ 
would have no node.  

For instance, let us consider that $W$ is given by  
\begin{equation}
W=-2\tanh((\eta-\xi)/\Delta),
\label{tanh}
\end{equation} 
where the parameter $\xi$ indicates the position of the bubble wall.
Then $U$ becomes 
\begin{equation}
 \left({2\over\Delta}-4\right){1\over\cosh^2(\eta-\xi)/\Delta}. 
\label{Uex}
\end{equation}
When $\Delta>1/2$, the potential $U$ becomes negative. 
Then $q$ has a node, and hence there are supercurvature modes. 
On the other hand, when $\Delta<1/2$, the potential $U$ is positive
and there are no supercurvature modes. 
Of course these results are consistent with the sufficient condition 
(\ref{suff}), which for the present case can be shown to imply
$\Delta>3/4$. Also, it is known that for the ``superpotential''
(\ref{tanh}) the SWKB condition is exact, so (\ref{swkbcond}) is
equivalent to the condition $\Delta>1/2$. 
Note that $\Delta \sim 1$
separates the thin and thick wall regimes, since for $\Delta\ll 1$ 
we may write 
$$
 \hbox{wall thickness}\sim a(\xi)\Delta \ll a(\xi)\sim 
  \hbox{wall radius}. 
$$
The form of the potential $V(\phi)$ corresponding to (3.19)
cannot be given in closed form for arbitrary $\Delta$. However, as
noted in \cite{supcrit}, for the case $\Delta=1$ it corresponds to the 
inverted quartic double well potential, which is discussed in the 
next subsection.
 
Here we note that the above method for constructing a model
does not prove that $\phi(\eta)$ will actually represent a 
relevant instanton, since for the same potential $V$ there may be 
another solution which gives a larger decay rate. This issue is 
discussed in \cite{supcrit}.

\subsection{Generalized Fubini instanton and the implications of 
supercurvature modes}
\label{subBB}

In flat space, the inverted quartic potential
\begin{equation}
V_f=-(\lambda/4) \phi^4
\label{phi4}
\end{equation} 
has the following solution, 
known as the Fubini instanton \cite{fubini,lewe}:
\begin{equation}
\phi_{flat}=\sqrt{8\over \lambda} {\rho \over r^2+\rho^2}
\label{fubini}
\end{equation}
Here, $r$ is the distance to the origin of polar coordinates,
and the arbitrary parameter $\rho$ can be interpreted as the 
size of the instanton. This is therefore a one-parameter 
family of instanton solutions which interpolate between the ``top'' 
of the potential $\phi=0$ at $r\to \infty$ (which can be considered 
as ``false vacuum'') and an 
arbitrary $\phi=(8/\lambda)^{1/2} \rho^{-1}$ down the hill at $r\to 0$
(which can be thought as true vacuum). Note that in this flat space model
there is no classical barrier between those two values of the field, and
hence the instanton represents tunneling without barriers \cite{lewe}.
The Euclidean action for 
(\ref{fubini}) is 
\begin{equation}
S_E={8\pi^2\over 3\lambda}.
\label{actionfubini}
\end{equation}
The fact that the solution exists for all sizes $\rho$ and that moreover
the action is independent of $\rho$ is a consequence of the conformal 
invariance of the potential (\ref{phi4}), which contains no mass scales.

Since de Sitter space is conformally flat, we can generalize the Fubini
instanton to de Sitter space. Therefore, for simplicity, 
we shall neglect the gravity of the bubble and we shall work directly 
on the 4-sphere. The metric can be written as a conformal factor times
the metric of flat space
\begin{equation}
ds^2= \Lambda^2 ds^2_{flat}= \Lambda^2 [dr^2+ r^2 dS_3],
\end{equation}
where 
\begin{equation}
\Lambda(Hr)={1\over 1+(Hr/2)^2},
\end{equation}
$H^{-1}$ is the radius of the 4-sphere and $dS_3$ is the metric on the
3-sphere.

Under conformal transformations
\begin{equation}
ds^2_{flat} \to ds^2=\Lambda^2 ds^2_{flat},\quad
\phi_{flat} \to \phi=\Lambda^{-1} \phi_{flat}
\label{phiconf}
\end{equation}
the Klein-Gordon operator has the following property 
(see e.g. \cite{bida})
$$
[\Box - 2H^2] \phi = \Lambda^{-3} \Box_{flat} \phi_{flat}.
$$
It follows that 
\begin{equation}
\phi=\Lambda^{-1} \phi_{flat}=\sqrt{2\over \lambda}\, {\rho\over 2} 
\,{4+H^2r^2 \over \rho^2+r^2}
\label{fubinidesitter}
\end{equation}
is a solution of the field equation $ \Box \phi = \partial V/\partial \phi$
for the new potential 
\begin{equation}
V= 2 H^2 \phi^2 - {\lambda\over 4} \phi^4 +\text{const.} \equiv -
{\lambda\over 4}(\phi^2-v^2)^2,
\label{newpotential}
\end{equation}
where
$$
v^2={2H^2\over \lambda}.
$$
Therefore, the solution (\ref{fubinidesitter}) generalizes the
Fubini instanton to de Sitter space. 

Note that we have added a conformal mass term to the potential $V(\phi)$.
The effect of this mass term is that now the potential looks like an
inverted double well, with a local
minimum at $\phi=0$. The solution (\ref{fubinidesitter}) now interpolates 
between
\begin{equation}
\phi_N:= v \bar\rho^{-1}
\label{phinucl}
\end{equation}
at the ``pole'' (or nucleation point) $r=0$ and 
$$
\phi_A:= v \bar\rho
$$
at the ``antipole'' (or antipodal point) $r\to \infty$.
Here we have introduced the parameter 
$\bar\rho:={\rho H/2}.$
The generalized instanton no longer represents tunneling
without barriers. It interpolates between a point inside the 
local well near $\phi=0$ and a point outside this local well,
beyond the local maxima at $\phi=v$.
Therefore this is a Coleman-De~Luccia instanton of the usual sort.
For $\bar\rho =1$, both turning points coincide
and we have $\phi_N=\phi_A=v$, so the Coleman-De~Luccia
instanton degenerates into a Hawking-Moss instanton, where the
field is constant throughout the sphere.

Introducing the variable $\eta = -\ln(rH/2)$ the
metric takes the form (\ref{metriceuclidean}) and the instanton 
(\ref{fubinidesitter}) takes the form
\begin{equation}
\phi_{\xi}(\eta)=v{ \cosh \eta \over \cosh(\eta-\xi)}.
\label{mtfubini}
\end{equation}
Here we have introduced a positive parameter $\xi$ which
can be interpreted as the position of the bubble wall in the conformal
coordinate $\eta$. It is related to $\bar\rho$ through
$$
\bar\rho=e^{-\xi}.
$$
The solution in the form (\ref{mtfubini}) was given
previously in \cite{supcrit}, where it was found using the method
described in subsection \ref{subA}. Indeed, it corresponds to the case
$\Delta=1$, which as mentioned there has a supercurvature mode.

Let us now discuss the dynamics of the resulting bubble. Because 
of conformal invariance, the value of the Euclidean action in curved space 
is also independent 
of the value of $\bar\rho$. Naively, one would then expect
different types of bubbles to nucleate with equal probability, 
characterized by a different initial value from which the field starts 
rolling down. 
However, as we shall see, this picture is too simplistic. 
What actually happens is that fluctuations of the supercurvature mode cause
different regions inside the {\em same} bubble to start from a different
value of $\phi_N$, so that all possibilities are sampled within just one 
bubble.

Let us recall the form of the 
supercurvature mode in this model.
Since (\ref{mtfubini}) is a solution for any value of
the parameter $\xi$, the associated zero mode
\begin{equation}
\varphi_{\xi}= {\partial \phi_{\xi}(\eta) \over \partial \xi},
\label{zeromode}
\end{equation}
must be a solution of the linearized equations of motion about the 
solution (\ref{mtfubini}). This perturbation only depends  
on $\eta$, and not on the spatial coordinates on the hyperboloid.
Hence, it must correspond to a solution of (\ref{ewq2})
with $p^2=-1=-\Lambda^2, l=m=0$, since only in this case the harmonic 
${\cal Y}^{\Lambda,00}$ in the expansion (\ref{eim}) is constant
[see Eq. (\ref{laplace})].
Indeed, 
\begin{equation}
{\bbox \chi}^{p^2=-1} \propto a(\eta) \varphi_0 \propto 
{-\sinh(\eta-\xi) \over \cosh^2(\eta-\xi)}
\label{zeromodexi}
\end{equation}
solves Eq. (\ref{ewq2}) and it is, moreover, normalizable in the
Schr{\"o}dinger
sense. However, this mode is real and hence it is not 
normalizable in the Klein-Gordon sense (its Klein-Gordon norm is zero). 
The reason is that its expected 
amplitude is actually infinite. Physically, this corresponds to the fact 
that all values of $\xi$  (that is, all values 
of the field $\phi_N$ at the time of nucleation) 
are equally probable.
To handle this case, the zero mode has to be quantized
in terms of ``position'' and 
``momentum'' operators associated to the collective coordinate $\xi$, 
and not in terms of creation and annihilation operators accompanied by 
the corresponding Klein-Gordon normalized modes. For
our purposes, however, it will be more instructive to break the
degeneracy by perturbing the
potential (\ref{newpotential}), so that not all values
of $\phi_N$ are equally probable.

Following \cite{supcrit}, we add the term 
\begin{equation}
\Delta V= \epsilon V_1(\phi),
\label{deltav}
\end{equation}
to the potential (\ref{newpotential}), where $\epsilon$ is a small
parameter. In this case, the Euclidean action evaluated on the field 
configuration $\phi=\phi_{\xi}+\Delta\phi$, where $\Delta\phi$ is a small 
perturbation of $O(\epsilon)$ which is otherwise arbitrary, is given by
\begin{equation}
S(\xi):=\Delta S_E[\phi_{\xi}+\Delta\phi]
= S_E^{0}[\phi_\xi+\Delta\phi]+
2\pi^2 \epsilon \int d\eta a^4 V_1(\phi_{\xi})+O(\epsilon^2).
\label{rhs}
\end{equation}
Since $\phi_{\xi}$ is a solution of the unperturbed 
potential, the field perturbation $\Delta\phi$ does not appear in the
right hand side of this equation to linear order. Thus, to the lowest
order in $\epsilon$, the instanton is still given by
(\ref{fubinidesitter}), but now with the value of $\xi=\xi^0$ that
minimizes the right hand side of (\ref{rhs}). 
This selects a ``preferred'' value $\phi_N^0$. Using (\ref{mtfubini}), 
one can change the variable of integration in (\ref{rhs}) from $\eta$ to 
$\phi_{\xi}$. We have
$$
S(\xi)= S_E^{0}+ {16\pi^2 \over H^4} 
{\bar\rho^2 \over (1-\bar\rho^2)^3} 
\epsilon \int_{\bar\rho^{-1}}^{\bar\rho}
(\bar\rho - \bar\phi) (1-\bar\rho\bar\phi) \bar\phi^{-4} V_1(\phi) d\bar\phi,
$$
where $\bar\phi=\phi/v$ is the dimensionless field. For definiteness, let
us take $V_1= H^4 \bar\phi^4 (\alpha_0 + \alpha_1 \bar\phi + 
\alpha_2\bar\phi^2)$. In this case we have
$$
S(\bar\rho)
=4 \pi^2 \epsilon \left[{\alpha_1\over 3} {1+\bar\rho^2\over \bar\rho}+
{\alpha_2\over 5}{1+\bar\rho^4 \over \bar\rho^2}\right]+\text{const.}
$$
This expression has extrema at
\begin{equation}
\bar\rho=\bar\rho_0=\beta \pm (\beta^2-1)^{1/2}
\label{beta}
\end{equation}
where $\beta=-5\alpha_1/12\alpha_2$. Of course, (\ref{beta}) is only valid
for $\beta>1$. For $\beta\leq 1$ the extremum is at $\bar\rho=1$ and
the Hawking-Moss transition is the dominant channel. The double sign in
(\ref{beta}) corresponds to the two different turning points for 
a given instanton, so actually there is just one extremum for each 
value of $\beta$.

One can estimate the spread in $\phi_N$ by
using the formula for the probability of nucleation
\begin{equation}
P(\phi_N) \sim e^{-S_E(\phi_N)}
\label{adiabatic}
\end{equation}
where the exponent is calculated by substituting the function 
$\phi_{\xi}$ which corresponds to a given value $\phi_N$ in
(\ref{rhs}). We have
$$
S_E=S[\xi^0]+{1\over 2} {d^2 S\over d\xi^2}(\xi-\xi^0)^2.
$$
Using $d\phi_N=\phi_N d\xi$, the variance of $\phi_N$ is 
given by
\begin{equation}
\langle(\Delta\phi_N)^2\rangle = (\phi_N^0)^2 
\left({d^2 S\over d\xi^2}\right)^{-1}.
\label{spread}
\end{equation}
In \cite{supcrit} it was shown that by adding the perturbation
(\ref{deltav}), the eigenvalue at $p^2=-1$ shifts to a new 
eigenvalue 
$$
\gamma:= p^2+1={3 H^2 \over 4\pi^2 v^2}{d^2 S\over d\xi^2}.
$$
Introducing this expression back in (\ref{spread}) we have
\begin{equation}
\langle(\Delta\phi_N)^2\rangle = {3 \rho_0^{-2} \over \pi^2 \gamma}.
\label{spreadn} 
\end{equation}
Here $\rho_0 =2\bar\rho_0 H^{-1}$ is the size of the bubble.  Note
that the eigenvalue $\gamma$ is of order $\epsilon$, and so as
expected the spread goes to infinity as we switch off the
perturbation.  Although the result (\ref{spreadn}) has been obtained
here rather heuristically from (\ref{adiabatic}), it can be justified
more rigorously by studying the r.m.s. fluctuation of $\phi$ in the
O(3,1) invariant quantum state, as it was done in \cite{modelq} for
two field models. In particular, by calculating the two point function
\cite{modelq}, it can be shown that the spatial correlations in $\phi$
at the time of nucleation decay on the co-moving scale $r \sim
\gamma^{-1}$, where $r\sim 1$ corresponds to the curvature scale on
the open sections. Therefore, we have the picture that inside of each
bubble there is an ensemble of regions of size $r\sim\gamma^{-1}$
where the value of $\phi_N$ is coherent. But the different regions of
the ensemble take all possible values of $\phi_N$, with statistical
average given by the peak value $\phi_N^0$ and with r.m.s. given by
(\ref{spread}). The situation here can be contrasted with the one for
the two-field models considered in \cite{modelq}. There, the mean
value of the slow roll field $\phi$ selected by the instanton solution
is actually zero, at the bottom of the slow roll potential, and
inflation is only due to localized fluctuations with r.m.s. given by
the quantum state. Here, the mean value of $\phi$ is nonzero, at a
certain value $\phi_N^0$ from which the field starts rolling down.

Also, one can show that r.m.s. amplitude
of the supercurvature modes with $l>0$ is of order $\rho_0^{-1}$, and
not of order $H$ as it is for subcurvature modes. This is a rather
general conclusion, and physically it corresponds
to the fact that, since these modes ``live'' in the bubble wall, they are 
excited due to the bubble acceleration, and not just due to the expansion
of the universe. To avoid too large CMB anisotropies caused by the
supercurvature modes (note that $\rho_0^{-1}>H$), in one field models we
must `tune'' the parameters to some extent, so that the radius of the
bubble is comparable to the Hubble radius. In our case, this is achieved
for $\beta \sim 1$. 

Unfortunately, the inverted quartic potential
is not too appropriate from the point of view of slow roll inflation. 
After nucleation the field would roll down to infinity in a finite 
proper time and the kinetic energy would soon dominate over the potential. 
Nevertheless, it appears that in any model with supercurvature modes
the picture described above should be qualitatively similar. The 
different values of $\phi$ at the time of nucleation would then result
in different lengths of the slow roll inflation, leading to an ensemble
of regions with different values of the density parameter. This is the
same picture that arises in the case of two field models studied in
\cite{modelq,anth}. More general single-field models with supercurvature
modes may be viable phenomenologically, although they are likely to be
rather fine tuned. This issue is currently under investigation. 

\section{Spectrum of scalar-type perturbation}
\label{specscal}
\subsection{The quantum state at $\eta_R=-\infty$}
\label{secsc0}

In our previous paper \cite{prev}, we showed that the general features
of the spectrum of the tensor-type perturbation can be understood
without specifying details of the potential model. This fact will be
briefly reviewed in section\ref{tenspec}.  Here we show that an
analogous argument also holds for the scalar type perturbation.  {}For
this purpose, it is convenient to use the
Eq.~(\ref{qeq}).  We expand the quantum operator $\hat{\bbox q}$
corresponding to the variable ${\bbox q}$ as
\begin{equation}
  \hat{\bbox q}=\sum \hat a^{(s)}_{p\ell m} {\cal N}^S_p 
             {\bbox q}^p (\eta) Y^{p\ell m}(\chi,\Omega) + 
  \sum \hat a^{(s)}_{\Lambda,\ell m} {\cal N}^S_{,\Lambda} 
  {\bbox q}^{,\Lambda}(\eta){\cal Y}^{\Lambda,\ell m}
(\chi,\Omega)+(\hbox{h.c.}). 
\end{equation} 
where $\hat a^{(s)}_{p\ell m}$, $\hat a^{(s)}_{\Lambda,\ell m}$ are the
annihilation operators.  Below, we shall analyse the behaviour of these
modes in region $C$.  The language of one dimensional scattering off a
potential barrier will be used. Then we will analytically continue the
modes inside region $R$ and compute its evolution till the end of
inflation.

Noting the limiting behaviors $\phi'\propto e^{\mp 2\eta_C}$ at
$\eta_C\to \pm \infty$, the potential term of Eq.~(\ref{qeq})
tends to
\begin{equation}
  {\kappa\over 2}{\phi'}^2+ 
      \phi'\left({1\over \phi'}\right)''
 \rightarrow 4 + O(a^2)\quad(\eta_C\rightarrow\pm\infty),
\end{equation}
and Eq.~(\ref{qeq}) reduces to 
\begin{equation}
 -{d^2\over d\eta_C^2} {\bbox q}^p=p^2 {\bbox q}^p
\quad(\eta_C\rightarrow\pm\infty). 
\label{nnn}
\end{equation}
Hence, for the continuous spectra one can choose the plane waves
$e^{\mp ip\eta_C}$ incident from $\eta_C=\pm\infty$ as the two
orthogonal solutions for $\bq_{\pm}^{p}$. These incident plane waves
interact with the potential and produce waves reflected back to
$\eta_C\to\pm\infty$ with reflection amplitudes $\varrho_{\pm}$, and
waves transmitted to $\eta_C\to\mp\infty$ with transmission amplitudes
$\sigma_{\pm}$. That is, ${\bbox q}_{\pm}^{p}$ in the limit
$\eta_C\to\pm\infty$ are given by
\begin{equation}
 i{\bbox q}_{+}^{p}= \left\{\begin{array}{ll} 
        \varrho_{+} \,e^{ip\eta_C} + e^{-ip\eta_C}, 
      &(\eta_C \rightarrow +\infty),\\ 
       \sigma_{+} \,e^{-ip\eta_C}, 
      &(\eta_C \rightarrow -\infty), 
\end{array}\right.
\label{bfqp}
\end{equation}
and
\begin{equation}
i{\bbox q}_{-}^{p} =\left\{\begin{array}{ll} 
       \sigma_{-} \,e^{ip\eta_C},
      &(\eta_C \rightarrow +\infty),\\
     \varrho_{-} \,e^{-ip\eta_C} + e^{ip\eta_C},
      &(\eta_C \rightarrow -\infty).
    \end{array}\right.
\label{bfqm}
\end{equation} 
Here we note that $\bq_+^{-p}$ and $\bq_-^{-p}$ 
are not independent of $\bq_+^{p}$ and $\bq_-^{p}$. Hence we restrict
$p$ to be non-negative. The reflection and transmission coefficients
satisfy the relations
\begin{eqnarray}
  && \vert\sigma_{+}  \vert^2
    =1-\vert\varrho_{+} \vert^2, 
\quad 
  \vert\sigma_{-}  \vert^2
    =1-\vert\varrho_{-} \vert^2, 
\label{Wron1}
\\ 
  && \sigma_{-} =\sigma_{+},
\quad \sigma_{+}  \bar\varrho_{-} +
\bar\sigma_{-}  \varrho_{+} = 0, 
\label{Wron2}
\end{eqnarray}
which follow from the fact that the Wronskian of any two solutions of 
(\ref{nnn}) is constant.
It is easy to see that
$\int_{-\infty}^{\infty}
d\eta\,\bq^p_{\sigma}\overline{\bq^{p'}_{\sigma'}}
=2\pi\delta(p-p')\delta_{\sigma\sigma'}$. 
Thus, from Eq.~(\ref{normalS}), the normalization constant 
${\cal N}^S_p$ is determined as 
\begin{equation}
\label{normcont}
 {\cal N}_p^S=\left(4p(p^2+4)\sinh\pi p\right)^{-1/2}. 
\end{equation}
The analytic continuation to region $R$ of Eqs. (\ref{bfqp}) and
(\ref{bfqm}) gives the continuous modes in the limit
$\eta_R\to-\infty$ as
\begin{eqnarray}
 {\bbox q}^p_{R+} & = & e^{\pi p/2} \varrho_{+} e^{-ip\eta_R} 
           + e^{-\pi p/2} e^{ip\eta_R}, \cr
 {\bbox q}^p_{R-} & = & e^{\pi p/2} \sigma_{-} e^{-ip\eta_R}, 
\label{chipR}
\end{eqnarray}
with the constraint on the coefficients, 
$|\varrho_{+}|^2+|\sigma_{-}|^2=1$. These play the role of the positive
frequency mode functions in region $R$.

For the discrete part of the spectrum, the normalizable solutions
${\bbox q}^{,\Lambda}$ can be choosen to have the form 
\begin{equation}
\label{discreteq}
{\bbox q}^{,\Lambda}=e^{\pm\Lambda\eta_C}\quad (\eta\to\mp\infty)\,.
\end{equation}
The normalization constant ${\cal N}^S_{,\Lambda}$ is determined from Eq.
(\ref{normsupB}),
\begin{equation}
\label{normsup}
{\cal N}_{,\Lambda}^S=\left(({4-\Lambda^2})\int_{-\infty}^{\infty}
    d\eta\, |\bq^{,\Lambda}|^2\right)^{-1/2}.
\end{equation}
The analytic continuation to region $R$ of Eq. (\ref{discreteq}) gives
the discrete modes in the limit $\eta_R\to-\infty$ as
\begin{equation}
{\bbox q}^{,\Lambda} = e^{i\Lambda\pi/2}e^{\Lambda\eta_R}.
\end{equation}

\subsection{General formula for the power spectrum}

The important quantity that determines the primordial density
perturbation spectrum as well as the large angle CMB anisotropies is the
curvature perturbation on the comoving hypersurface, ${\cal R}_c$. 
The comoving hypersurface is the one on which the scalar field
fluctuation $\varphi$ vanishes. In the present case, ${\cal R}_c$ is
given by
\begin{equation}
  \label{calRc}
  {\cal R}_c^p=-{\cal N}_p^S \left({{\kappa}\phi'\bq^p\over2a}
    +{{\cal H}\over \phi'}\varphi_N^p\right).
\end{equation}
Using Eq. (\ref{dotP}), it can be written in terms of ${\bbox q}$ alone,
\begin{equation}
\label{calRq}
{\cal R}_c^p = -{\cal N}^S_p\left[\frac{\kappa}{2}
    \frac{\phi'}{a}+
  \frac{{\cal H}}{a\phi'{}^2}\frac{d}{d\eta_R}\phi'\right]{\bbox q}^p.
\end{equation} 
The evolution of the scalar-type perturbation is governed by
the analytic continuation of Eq.~(\ref{qeq}) to region $R$.
It takes exactly the same form as the one in region $C$:
\begin{equation} 
\left[\frac{d^2}{d\eta_R^2}+(p^2+4)-\phi'\left({1\over \phi'}\right)''
        -{\kappa\over 2}{\phi'}^2 \right]{\bq}^p=0.
\label{qeqopen}
\end{equation}
Now, we will define a function that will gives us the value at the
end of inflation of the curvature perturbation ${\cal R}_c^p$
which is generated by a scalar perturbation ${\bbox q}^p$ having the form
$e^{ip\eta_R}$ on the lightcone when it enters inside region $R$ (i.e.
when $\eta_R\to-\infty$). From Eq. (\ref{calRq}), this ``transfer
function'' for the scalar curvature perturbation is given by
\begin{equation}
{\cal T}^p_S := \lim_{\eta_R\to \eta_{\rm end}} \left[-\frac{\kappa}{2}
  \frac{\phi'}{a}-
\frac{{\cal H}}{a\phi'{}^2}\frac{d}{d\eta_R} \phi'\right]
  {\bbox{\tilde q}}^p,
\label{transfersc}
\end{equation}
where, accordingly, ${\bbox{\tilde q}}^p$ is the solution of Eq.
(\ref{qeqopen}) which behaves as ${\bbox{\tilde q}}^p\to
e^{ip\eta_R}$ when $\eta_R\to-\infty$, and where we have drop the
normalization constant ${\cal N}_p^S$. Here $\eta_{\rm end}$ is the
value of the conformal time $\eta_R$ for which $a\to\infty$. Taking
into account that the quantum state for ${\bbox q}^p$ inside region $R$
is given by Eqs.  (\ref{chipR}), it is easily seen that the continuous
part of the primordial spectrum for ${\cal R}_c$ at the ``end of
inflation'' can be computed from
\begin{eqnarray}
  \langle|{\cal R}^p_c|^2\rangle &=& |{\cal R}_c^{p+}|^2 +  |{\cal
      R}_c^{p-}|^2 =
2|{\cal N}^S_p|^2|{\cal T}^p_S|^2
  \left(\cosh \pi p+\Re\left(\varrho_+\frac{\overline{{\cal T}_S^p}}
      {{\cal T}_S^p} \right)\right)\,. \nonumber\\
\end{eqnarray}
Using Eq. (\ref{normcont}) we have
\begin{equation}
  \langle|{\cal R}^p_c|^2\rangle = 
  \frac{|{\cal T}_S^p|^2}{2p (p^2+4)\sinh \pi p}
  \left(\cosh \pi p+\Re\left(\varrho_+\frac{\overline{{\cal T}_S^p}}
      {{\cal T}_S^p} \right)\right).
\label{general}
\end{equation}

For the discrete modes we define the transfer function as follows,
\begin{equation}
{\cal T}^{,\Lambda}_S := \lim_{\eta_R\to\eta_{\rm end}}\left[-\frac{\kappa}{2}
  \frac{\phi'}{a}-
  \frac{{\cal H}}{a\phi'{}^2}\frac{d}{d\eta_R} \phi'\right] 
 {\bbox{\tilde q}}^{,\Lambda},
\end{equation}  
where ${\bbox{\tilde q}}^{,\Lambda}$ is the solution of Eq.
(\ref{qeqopen}) which tends to $e^{\Lambda\eta_R}$ on the lightcone
inside region $R$. The discrete part of the primordial spectrum for the
curvature perturbation is then given by
\begin{equation}
  \langle|{\cal R}^{,\Lambda}_c|^2\rangle = |{\cal
  R}_{c}^{,\Lambda}|^2 =  
  |{\cal N}^S_{,\Lambda}|^2|{\cal T}^{,\Lambda}_S|^2\,,
\label{generalsup}
\end{equation}
where ${\cal N}_S^{,\Lambda}$ is given by Eq. (\ref{normsup}).

Thus, in order to find the power spectrum for scalar-type
perturbations one has to solve two separate problems. First, for the
continuous spectra, one has to solve the scattering problem posed in
the previous subsection in order to find the reflection coefficient
$\rho_+$.  Roughly speaking, this accounts for the scattering of
scalar modes off the bubble wall, which determines the initial quantum
state for these modes at the beginning of the second stage of
inflation inside the bubble. For the discrete spectrum, one has to
find the normalization of the modes, which will give us the ``size''
of their effect. Then, one has to compute the transfer function ${\cal
  T}_S$, which accounts for the evolution of the modes during the
second stage of inflation.

The expression (\ref{general}) is of general validity. Typically,
however, both the reflection coefficient and transfer function can
only be found numerically. In the appendices we discuss some exactly
soluble models, but for the rest of this section we shall consider the
weak backreaction case, which can also be treated analytically.

\subsection{Weak backreaction case}

We now consider the evolution of the scalar-type perturbation inside the 
bubble in the weak backreaction case. For notational simplicity, we set 
$t=\tau_R$ and omit the
suffix $R$ from all the variables in the rest of this subsection
unless ambiguity arises.

In general,
one cannot assume the slow roll motion of $\phi$
immediately after nucleation. Nevertheless, it can be assumed that 
the back reaction of the motion of $\phi$ to the cosmic expansion
is negligible at this stage when the universe is curvature-dominated.
Then, after a few Hubble expansion times, the universe starts to expand
exponentially and the slow roll assumption becomes valid.
Hence it is fair to assume
\begin{equation}
\kappa\dot\phi^2\ll {{\cal H}^2\over a^2}\approx H^2+{1\over a^{2}},
\label{condidotphi}
\end{equation} 
where $H^2:=\kappa V/3$. We call this the small back reaction
approximation, since it corresponds to the limit in which the back
reaction of the inflaton field dynamics to the geometry is small.
In this limit, $a$ can be approximated by that of de Sitter space
as long as the time duration considered, $\Delta t$, is not much 
greater than $H^{-1}$. Also, we shall make the approximation that 
$\partial^2 V$ varies slowly compared with the expansion time.
Furthermore, since the initial condition and the evolution of the mode
function $\bq^p$ for $p\gg1$ will be the same as the case of the flat
universe inflation, we focus our attention on the modes with
$p$ not too greater than unity.

%The evolution of the scalar-type perturbation is governed by
%the analytic continuation of Eq.~(\ref{qeq}) to region $R$.
%It takes exactly the same form as the one in region $C$:
%\begin{equation} 
%{\bq^p}''+\left[p^2+4-\phi'\left({1\over \phi'}\right)''
%        -{\kappa\over 2}{\phi'}^2 \right]{\bq}^p=0.
%\label{qeqopen}
%\end{equation}

To discuss the evolution in the small back reaction limit,
we find it is more convenient to use the equation for the scalar field
perturbation in the Newton gauge, Eq.~(\ref{phieqnew}), instead of
working directly with (\ref{qeqopen}). 
Setting 
\begin{equation}
 \varphi_N^p ={\chi^p\over a}\,,  
\end{equation}
rewriting Eq.~(\ref{phieqnew}) in
terms of $\chi^p$ and $\eta_R$, and performing analytic continuation,
we obtain
\begin{equation}
  \label{chieq}
{\chi^p}''+\left[p^2+2(1-{\cal H}^2)+\partial^2V\,a^2\right]\chi^p
 ={5\over2}\kappa{\phi'}^2\chi^p
   - \kappa {\phi'}^2{\bq^p}'\,.
\end{equation}
Also from Eq.~(\ref{dotP}), $\bq^p$ and $\chi^p$ are related as
\begin{equation}
  \label{bqchi}
  \chi^p={1\over \phi'}(\phi'{\bbox q}^p)'= {\bq^p}'+{\phi''\over\phi'}\bq^p
\end{equation}
Note that this is basically the same as Eq.(\ref{susytrans2}), except
that here we are using the expansion (\ref{qexpa}) for ${\bbox q}$ and
for ${\chi}$, with the same normalization constants ${\cal
  N}^S_p$, whereas in (\ref{susytrans2}) we were using the
Schr{\"o}dinger normalization for both sets of eigenstates. In the
limit $\eta\to-\infty$, Eq. (\ref{bqchi}) reduces to
\begin{equation}
\chi^p \to \left({d\over d\eta}+2\right) {\bbox q}^p \hskip 1cm 
(\eta\to -\infty).
\label{bqchia}
\end{equation}

{}From the above equation (\ref{bqchi}) and the asymptotic behavior 
of $\bq^p$ and
$\phi'$ at $a\to0$ ($\eta\to-\infty$), namely $\bq^p\sim e^{\pm ip\eta}$
and $\phi'\sim e^{2\eta}$, we find ${\bq^p}'\sim\chi^p$ or
smaller for $a\ll1$. On the other 
hand, the behavior of $\bq^p$ for $a\gg1$ can be deduced from
Eq.~(\ref{qeqopen}). Under the slow roll approximation, we find
\begin{equation}
  \label{qasympt}
  \bq^p\sim e^{\int\delta Hdt}\,,\quad e^{-\int(1+\delta)Hdt}\,;
\quad \delta:={3\over2}\kappa{\dot\phi^2\over H^2}
-{1\over3}{\partial^2V\over H^2}\,.
\end{equation}
Hence, assuming the growing mode (the first solution in the above)
dominates, we have ${\bq^p}'\ll (\phi''/\phi')\bq^p\sim\chi^p$. Then,
since ${\cal H}^2-1\approx H^2a^2$ and
$\kappa{\phi'}^2=\kappa\dot\phi^2a^2$, we may neglect the right hand
side of Eq.~(\ref{chieq}) in the small back reaction approximation.

Also, for the stage $(Ha)^2\gg1$, we approximately have
\begin{equation}
{\cal R}_c^p=-{\cal N}_p^S {H\over{a \dot\phi}}\chi^p\,.
\end{equation}
Just as in the case of the flat universe inflation, this turns out to
be constant in time for $H^2a^2\gg p^2$.  For definiteness, let us
explicitly show this fact and derive the power spectrum of ${\cal
  R}_c$.

The transfer function defined in Eq. (\ref{transfersc}) is now
approximately given by
\begin{equation}
{\cal T}^p_S \approx \lim_{\eta \to 0}\left[-\frac{H}{a\dot \phi} 
  \widetilde\chi^p \right],
\end{equation}
where, using Eq. (\ref{bqchia}), $\widetilde\chi^p$ is the solution of Eq.
(\ref{chieq}) which behaves as $\widetilde\chi^p\to(2+ip)e^{i\eta}$ when
$\eta\to-\infty$. Now let us find the solution to Eq.
(\ref{chieq}) in the small back reaction limit. With the
approximations above, and assuming the time variation of
$M^2=\partial^2V$ is small compared with the expansion time scale, the
general solution to Eq. (\ref{chieq}) is
\begin{equation}
 \chi^p
=\alpha P^{ip}_{\nu'}(-\coth\eta)+\beta P^{-ip}_{\nu'}(-\coth\eta),
\label{chipsol}
\end{equation}
where $-\coth\eta=\cosh H t$, and $\nu'$ is given by
\begin{equation}
 \nu' :=\sqrt{{9\over 4}-{M^2\over H^2}}-{1\over 2}\,.
\end{equation}
It should be noted that there are cases in which our assumptions may
not be valid.  For example, in a model of one-bubble inflation
recently proposed by Linde \cite{Lindetoy}, the time variation of
$M^2$ is significant at the early stage of inflation inside the
bubble.

The approximate solution (\ref{chipsol}) is valid for $t\alt\hbox{a
  few}\times H^{-1}$, i.e., until $aH$ is not too much greater than
unity. For definiteness, we write this condition as $aH\ll B$ where
$B$ is a big number, say $\sim e^{10}$.  Noticing that
\begin{equation}
 P^{-ip}_{\nu'}(-\coth\eta)\to{ e^{ip\eta}\over \Gamma(ip+1)}
\quad(\eta\to-\infty), 
\end{equation} 
we find
\begin{equation}
\label{evolchi}
  \widetilde\chi^p = (2+ip)\Gamma(ip+1) 
              P^{-ip}_{\nu'}(-\coth\eta)\,.
\end{equation}
Let us consider the behavior of $\widetilde\chi^p$ at the stage when $1\ll
H^2a^2\ll B^2$. Note that, for $p=O(1)$, we have $p^2\ll H^2a^2$ as
well. Thus we consider the stage $1+p^2\ll H^2a^2\ll B^2$.
At this stage, the approximate solution (\ref{evolchi}) is still valid
and, using the assymptotic form of the Legendre function, the expression of 
$\widetilde\chi^p$ reduces to 
\begin{equation}
 \widetilde\chi^p\rightarrow {(2+ip)\over\sqrt{\pi}}
   {\Gamma(ip+1)\Gamma\left(\nu'+{1\over 2}\right)\over 
     \Gamma\left(\nu'+ip+1\right)}e^{\nu'Ht}.
\end{equation}

Now we consider the evolution of the background inflaton field at
the stage when $1+p^2\ll H^2a^2\ll B^2$. In accordance with our
assumption, the potential $V(\phi)$ during this epoch can be
approximated by
\begin{equation}
 V(\phi)=V_p+\partial V_p (\phi-\phi_p)
+{1\over 2}M^2(\phi-\phi_p)^2, 
\end{equation}
where $\phi_p$ is the value of $\phi$ at a fiducial time $t=t_p$.
A convenient choice of $t_p$ is the time at which
 $\hbox{a few}\,\times(1+p^2)=a^2H^2$, which is essentially the horizon
 crossing time.
During this stage, one can approximate the evolution equation of the 
background field as 
\begin{equation}
 \ddot\phi+3H\dot\phi=-\partial V_p-M^2(\phi-\phi_p),
\end{equation}
where $H=H(t_p)$. The solution is
\begin{equation}
\phi-\phi_p = -{\partial V_p\over M^2}
+c_1 e^{-\lambda_1 H(t-t_p)} 
        +c_2 e^{-\lambda_2 H(t-t_p)}, 
\end{equation}
where $c_1$ and $c_2$ are integration constants and 
\begin{equation}
 \lambda_1={3-\sqrt{9-4M^2/H^2}\over 2},\quad
 \lambda_2={3+\sqrt{9-4M^2/H^2}\over 2}.
\end{equation}
The terms with $c_1$ and $c_2$ correspond to 
the growing and decaying solutions, respectively. 
Then neglecting the decaying solution, we can 
evaluate $\dot\phi$ during this stage as
\begin{equation}
\dot \phi= {(-\partial V_p)\over H \lambda_2} e^{-\lambda_1H(t-t_p)}. 
\label{dotphi}
\end{equation}
Hence, noting that $\lambda_1=-\nu'+1$, we have $H/(a\dot\phi)\propto
e^{\nu'Ht}\propto\widetilde\chi^p$. Thus ${\cal R}_c^p$ stays constant
in time for $1+p^2\ll H^2a^2\ll B^2$. After this stage, the constancy
of ${\cal R}_c^p$ can be shown in the exactly the same manner as in
the case of the flat universe inflation.

Collecting the results above,  we find that the transfer function is
given by
\begin{equation}
{\cal T}_S^p=\left(\frac{H^3\lambda_2}
{(2 aH)^{\lambda_1}\partial V}\right)_{t=t_p} 
\frac{2(2+ip)\Gamma(\nu'+\frac{1}{2})\Gamma(ip+1)}
{\sqrt{\pi}\,\Gamma(\nu'+ip+1)}.
\end{equation}
Notice that the factor $(2aH)^{\lambda_1}\vert_{t=t_p}\approx 1$ for
models with $M^2<<H^2$. Finally, the power spectrum of ${\cal R}_c^p$
at the end of inflation is given by
\begin{equation} 
 \left\langle\vert {\cal R}_c^p\vert^2\right\rangle 
   = \left({H^3 \lambda_2
\over {(2 aH)^{\lambda_1}{\partial V}}}\right)^2_{t=t_p}
    {2(\cosh\pi p +\cos \delta_p)\over\sinh^2\pi p}
    \left\vert {\Gamma\left(\nu'+{1\over 2}\right)\over 
          \Gamma\left(\nu'+ip+1\right)}\right\vert^2,
\label{scspec}
\end{equation}
where $\delta_p$ is a phase defined by 
\begin{equation}
 e^{i \delta_p}:=
         {(2-ip)\Gamma(1-ip)\Gamma(\nu'-ip+1)\rho_+\over
          (2+ip)\Gamma(1+ip)\Gamma(\nu'+ip+1)|\rho_+|}\,. 
\end{equation}
If we set $M^2=0$, we recover the result presented in Eq.(4.2) 
of \cite{YST}.%{\bf PRD54}, p5031. 

As for the supercurvature modes, although we have argued in section
\ref{secsuper} that they are not expected in thin wall models, they
are not excluded in the general case. The normalization appearing in
Eq. (\ref{normsup}) and Eq. (\ref{generalsup}) is strongly model
dependent. As mentioned at the end of section \ref{secsuper},
supercurvature modes can impose severe constraints on the parameters
of the model. For a discussion of supercurvature modes in two-field
models see \cite{cansurvive,modelq}
 
In the above discussion, we have assumed the small time variation of
$M^2$ as well as the small back reaction limit.  However, as we have
noted, there are models in which the former condition is violated
\cite{Lindetoy}.  Furthermore, it is possible to consider a model in
which the back reaction is significant during the first few expansion
times inside the nucleated bubble. For such a model, the spectrum at
$p^2\alt1$ can be quite different from the one we derived above, and
one should resort to the general expression (\ref{general}), which
typically will require numerical evaluation of the reflection
coefficient and transfer function. Nevertheless, it is worthwhile to
note that the contribution of the $p\ll 1$ part of the scalar spectrum
to the large angle CMB anisotropies is negligible \cite{YST}. This is
due to the fact that $1/p^2$ factor in the normalization constant at
$p^2\to0$ is canceled by the factor $p^2$ coming from the harmonics
$|Y^{p\ell m}|^2$.

\section{tensor-type perturbation}

\subsection{Spectrum at $\eta_R\to -\infty$}
\label{tenspec}

In our previous paper \cite{prev}, we have shown that 
the spectrum of the tensor-type perturbation is constrained
to have a certain restricted form. 
Here we briefly review the result.
The argument is parallel to the one for the scalar-type perturbation
given in section \ref{secsc0}.

We expand the quantum operator $\hat{\bbox w}$ corresponding to 
the variable ${\bbox w}$ as 
\begin{equation}
  \hat{\bbox w}=\sum \hat a^{(t)}_{p\ell m} {\cal N}_{p\ell}^T 
             {\bbox w}^p (\eta_C) Y^{p\ell m} + (\hbox{h.c.}),
\end{equation} 
where $\bw^p$ satisfies the eigenvalue equation (\ref{weq}).
As noted in section \ref{actten}, there exists no supercurvature
mode in the tensor-type perturbation.
Note also that the potential term in Eq.~(\ref{weq}) goes to zero
at both boundaries of the region $C$. 
Hence, as the two orthogonal solutions ${\bbox w}_{(\pm)}^{p}$,
we may take those having the asymptotic behaviors as
\begin{equation}
 i{\bbox w}_{+}^{p}= \left\{\begin{array}{ll} 
        \varrho_{+} \,e^{ip\eta_C} + e^{-ip\eta_C}, 
      &(\eta_C \rightarrow +\infty),\\ 
       \sigma_{+} \,e^{-ip\eta_C}, 
      &(\eta_C \rightarrow -\infty), 
\end{array}\right.
\label{bfqpT}
\end{equation}
and
\begin{equation}
i{\bbox w}_{-}^{p} =\left\{\begin{array}{ll} 
       \sigma_{-} \,e^{ip\eta_C},
      &(\eta_C \rightarrow +\infty),\\
     \varrho_{-} \,e^{-ip\eta_C} + e^{ip\eta_C},
      &(\eta_C \rightarrow -\infty).
    \end{array}\right.
\label{bfqmT}
\end{equation} 
As in the case of scalar-type perturbation, using the Wronskian
relations we obtain
\begin{eqnarray}
  && \vert\sigma_{+}  \vert^2
    =1-\vert\varrho_{+} \vert^2, 
\quad 
  \vert\sigma_{-}  \vert^2
    =1-\vert\varrho_{-} \vert^2, 
\label{Wron1T}
\\ 
  && \sigma_{-} =\sigma_{+},
\quad \sigma_{+}  \bar\varrho_{-} +
\bar\sigma_{-}  \varrho_{+} = 0. 
\label{Wron2T}
\end{eqnarray}
We also have 
$\int_{-\infty}^{\infty} d\eta 
\bw^p_{\sigma}\overline{\bw^{p'}_{\sigma'}} 
=2\pi \delta(p-p')\delta_{\sigma\sigma'}$.
Thus, from Eq.~(\ref{normtilT}), the normalization constant 
$\tilde{\cal N}_p^T$ is determined as 
\begin{equation}
 \tilde{\cal N}_{p}^T= \sqrt{(p^2+1)\over \kappa p\sinh\pi p}. 
\label{tilcalNp}
\end{equation}

{}Finally the analytic continuation of the two independent modes 
to region $R$ gives
\begin{eqnarray}
 {\bw}^p_{R+} & = & e^{\pi p/2} \varrho_{+} e^{-ip\eta_R} 
           + e^{-\pi p/2} e^{ip\eta_R}, \cr
 {\bw}^p_{R-} & = & e^{\pi p/2} \sigma_{-} e^{-ip\eta_R}, 
\label{bwR}
\end{eqnarray}
with a constraint on coefficients, 
$|\varrho_{+}|^2+|\sigma_{-}|^2=1$. 
Thus we obtain
\begin{equation}
  \label{bwamp}
|\bw^p_{R+}|^2+|\bw^p_{R-}|^2=2(\cosh\pi p+\Re\rho_+),
\end{equation}
at $\eta_R\to-\infty$. This bound is obtained 
without specifying details of the model. For $p^2\gg1$,
 the spectrum at $\eta_R\to-\infty$ is
model-independent. On the other hand, the spectrum for small
 $p^2$ is sensitive to the choice of a model. It should be noted,
however, the reflection coefficient $\rho_+$ can be written in the form,
\begin{equation}
  \label{rhoplus}
  \rho_+=-|\rho_+(p)|e^{2ip\alpha(p)}
\end{equation}
where $|\rho_+(0)|=1$ and $\alpha(0)$ is finite.

As mentioned in the previous section, in the case of the scalar-type
perturbation, the contribution of the $p^2\ll 1$ modes to the CMB
anisotropy is negligible. Hence we did not have to worry about the
behavior of the spectrum at $p^2\ll 1$. On the contrary, in the case of
the tensor perturbation, the behavior of the low frequency spectrum 
does affect the large angle CMB anisotropy significantly.
This is because there is an extra factor of $1/p^2$ in the normalized
tensor harmonics $Y_{ij}^{(+)p\ell m}$ (see Eq.~(\ref{defYij})) as
compared with the scalar harmonics $Y^{p\ell m}$. In fact, if $\rho_+$
were not to have the behavior shown in Eq.~(\ref{rhoplus}), the CMB
anisotropy due to tensor-type perturbations would have diverged.
Hence, we must be more careful when dealing with the low frequency 
modes in the present case. In fact, the so-called wall fluctuation
modes, which give an important contribution to the CMB anisotropy,
are due to these modes \cite{STY}.  

\subsection{General formula for the power spectrum}

As in the case of scalar perturbations, we shall give a general
formula for the power spectrum of tensor modes in terms of the
reflection coefficient for the corresponding transfer function.

Let us first recapitulate the evolution
equation for the tensor perturbation. The equation for $\bw^p$ is
\begin{equation}
\left[{d^2\over d\eta_R^2}-{\kappa\over 2}{\phi'}^2+p^2\right]\bw^p=0. 
\label{tensoreq}
\end{equation}
Equivalently, the equation for $Q_T^p$ is
\begin{equation}
\left[{d^2\over d\tau_R^2}+3{\dot a\over a}{d\over d\tau_R}
+{p^2+1\over a^2} \right]{Q_T^p\over a^2}=0\,.
\label{QTeq2}
\end{equation}
The variable $\bw^p$ is expressed in terms of $Q_T^p$ 
as Eq.~(\ref{bwtoQT}). Conversely, substituting Eq.~(\ref{bwtoQT}) into
Eq.~(\ref{QTeq2}) yields the expression for $Q_T^p$ in terms of
$\bw^p$:
\begin{equation}
{Q_T^p\over a^2}=-{\kappa\over a(p^2+1)}{d\over d\tau_R}(a\bw^p).
\label{QTofbw}
\end{equation}

As for scalar perturbations, we will define an auxiliary ``transfer
function'' ${\cal T}_T$ in order to compute the power spectrum for the
variable $Q_T^p/a^2$ at the end of inflation . For tensor
perturbations it is defined as follows,
\begin{equation}
{\cal T}_T^p := \lim_{\eta_R\to\eta_{\rm end}} -\frac{\kappa}{a^2(p^2+1)}
\frac{d}{d\eta_R} a \tbw^p,
\end{equation}
where $\tbw^p$ is the solution of Eq. (\ref{tensoreq}) which behaves
as $\tbw^p\to e^{ip\eta_R}$ when $\eta_R\to-\infty$. Then, taking into
account the form of the quantum state on the lightcone inside region
$R$, Eq.  (\ref{bwR}), the power spectrum of $Q_T^p/a^2$ can be
expressed as
\begin{eqnarray}
  \langle|U_{p\ell m}^{(+)}|^2\rangle  = 
|\tilde {\cal N}^T_p |^2\left<\left|\frac{Q^p_T}{a^2}\right|^2\right> 
   = |\tilde {\cal N}^T_p |^2 \left(\left|\frac{Q^{p+}_{T}}{a^2}\right|^2+
     \left|\frac{Q^{p-}_{T}}{a^2}\right|^2\right)=
   2|\tilde{\cal N}^T_p|^2|{\cal T}_T^p|^2 
   \left(\cosh \pi p+\Re\left(\varrho_+\frac{\overline{{\cal T}_T^p}}
    {{\cal T}_T^p}\right)\right)\,.
\end{eqnarray}
Here, we have introduced the variable $U_{p\ell m}^{(+)}$ used in
\cite{TStensor,prev}, in terms of which the metric perturbations reads
\begin{equation}
\bh_{ij}[Q_T]=\sum U_{p\ell m}^{(+)} Y^{(+)p\ell m}_{ij}
\end{equation} 
Thus, from Eq. (\ref{defbh}), we have $U_{plm}^{(+)}=\tilde{\cal
  N}^T_p\,Q^p_T/a^2$.

Finally, the tensor power spectrum is given by
\begin{equation} 
 \langle|U_{p\ell m}^{(+)}|^2\rangle   = 2 
 \frac{|{\cal T}_T^p|^2 (p^2+1)}{\kappa p \sinh \pi p}
   \left(\cosh \pi p+\Re\left(\varrho_+\frac{\overline{{\cal T}_T^p}}
      {{\cal T}_T^p} \right)\right).
\label{generalT}
\end{equation}
In the appendices we present some cases where the transfer function
can be found analytically. However, for the remainder of this section
we turn attention to the weak backreaction case.

\subsection{Weak backreaction case}
\label{tensorin}

Again, as in the case of the scalar-type perturbation, we omit the
suffix $R$ from all the variables and set $t=\tau_R$ throughout this
subsection. We also adopt the small back reaction approximation, i.e.,
we assume $\kappa{\phi'}^2\ll{\cal H}^2$.  

%Let us first recapitulate the evolution
%equation for the tensor perturbation. The equation for $\bw^p$ is
%\begin{equation}
%\left[{d^2\over d\eta^2}-{\kappa\over 2}{\phi'}^2+p^2\right]\bw^p=0. 
%\label{tensoreq}
%\end{equation}
%Equivalently, the equation for $Q_T^p$ is
%\begin{equation}
%\left[{d^2\over d\tau^2}+3{\dot a\over a}{d\over d\tau}
%+{p^2+1\over a^2} \right]{Q_T^p\over a^2}=0\,.
%\label{QTeq2}
%\end{equation}
%The variable $\bw^p$ is expressed in terms of $Q_T^p$ 
%as Eq.~(\ref{bwtoQT}). Conversely, substituting Eq.~(\ref{bwtoQT}) into
%Eq.~(\ref{QTeq2}) yields the expression for $Q_T^p$ in terms of
%$\bw^p$:
%\begin{equation}
%{Q_T^p\over a^2}=-{\kappa\over a(p^2+1)}{d\over dt}(a\bw^p).
%\label{QTofbw}
%\end{equation}

At the first stage when $a^2\ll1$, we have $\bw^p\sim e^{\pm ip\eta}$.
In the small back reaction limit, this solution is valid as long as
the inequality $\kappa{\phi'}^2=\kappa\dot\phi^2a^2\ll p^2$
is satisfied. On the other hand, for $p^2+1\ll H^2a^2$,
Eq.~(\ref{QTeq2}) tells us that $Q_T/a^2$ becomes constant, save the
decaying solution that dies off as $\sim a^{-3}$.
Hence, for $p^2\agt1$, there exists a stage during which
$\kappa\dot\phi^2\ll p^2/a^2\ll H^2$ is satisfied. One can then evaluate
the final amplitude of $Q_T^p/a^2$ approximately at the $t=t_p$ when
$\hbox{a few}\,\times(p^2+1)=a^2H^2$ to obtain
\begin{equation}
{Q_T^p\over a^2}\approx-{\kappa\over (p^2+1)}(H\bw^p)_{t=t_p}.
\label{QTapprox}
\end{equation}
In this case, the transfer function greatly simplifies, being given by
\begin{equation}
{\cal T}_T^p \approx-{\kappa\over (p^2+1)}(H\tbw^p)_{t=t_p}.\label{traT}
\end{equation}
Calculating the spectrum of $Q_T^p/a^2$, we recover Eq.(4.8) of our
previous paper \cite{prev}, where the effect of the evolution inside
the bubble was neglected. Under this assumption, the transfer function
is readily given by ${\cal T}_T^p=-\kappa H/(p^2+1)$.

However, as already mentioned, the low frequency modes are very
important for the tensor-type perturbation. To evaluate the spectrum at
$p^2\alt 1$, we perform a perturbation analysis with respect to
$\kappa\phi'{}^2$.

First note that, once the universe is no longer exactly
de Sitter, the limit $\eta\to0$ does not correspond to $a\to\infty$ any
more. To accommodate this shift in $\eta$, we introduce a new time
coordinate $\zeta$ such that it satisfies $d\zeta=d\eta$ and approaches
zero for $a\to\infty$, and write the scale factor as 
\begin{equation}
 a={-1\over Q(\zeta)\sinh \zeta}\,,
\end{equation}
where the function $Q$ takes care of the deviation from the
exact de Sitter metric. It is assumed that $Q$ approaches a constant
at $\zeta\to-\infty$. 
Then we have
\begin{equation}
 {a'\over a}=-\coth\zeta-{Q'\over Q}\,,
\end{equation}
and the Friedmann equation becomes 
\begin{equation}
 h^2=Q^2\left[1+2\sinh\zeta\cosh\zeta{Q'\over Q}
   +\sinh^2\zeta\left({Q'\over Q}\right)^2 \right],
\label{htoQ}
\end{equation}
where the function $h$ is defined by
\begin{eqnarray}
 h^2(\zeta):={\kappa\over 3}\left({1\over 2}\dot\phi{}^2+ V\right).
\label{defh}
\end{eqnarray}
The time derivative of Eq.~(\ref{htoQ}) gives
\begin{equation}
 {\kappa\over 2}\phi'{}^2 =-\left({a'\over a}\right)'+a^2 h^2 
    =\left({Q'\over Q}\right)'+\left({Q'\over Q}\right)^2+2\coth\zeta 
     {Q'\over Q}. 
\label{Qeq}
\end{equation}
In the limit $\zeta\to-\infty$, we assume that the scale
factor approach that of the exact de Sitter universe. Hence we must
have
\begin{equation}
  \label{ato0lim}
  a\to{2e^{\eta}\over h_0}={2e^{\zeta}\over Q_0}\,, 
  \quad \zeta\to-\infty,,
\end{equation}
where $h_0=h(-\infty)$ and $Q_0=Q(-\infty)$. This gives
\begin{equation}
  h_0=Q_0e^{\eta-\zeta}=:Q_0e^{\Delta\eta}\,.
\end{equation}
Thus the correction to the conformal time, $\Delta\eta$, is
obtained by evaluating $Q_0$. Let us carry this out.

To the first order in $\kappa$, since $Q'/Q=O(\kappa{\phi'}^2)$, we can
neglect the quadratic term in Eq.~(\ref{Qeq}). 
Then it can be integrated to give
\begin{equation}
{Q'\over Q}={-1\over \sinh^2\zeta}
 \int_{\zeta}^0 d\zeta'\,{\kappa\over 2}{\phi'}^2 \sinh^2\zeta',
\end{equation}
where we have imposed the boundary condition that 
$\sinh^2\zeta Q'/Q\to0$ for $\zeta\to0$, which means the expansion is
almost de Sitter at late times. 
Inserting this to Eq.~(\ref{htoQ}) and taking
the limit $\zeta\to-\infty$, we obtain
\begin{equation}
e^{2\Delta\eta}
=1+\kappa\int_{-\infty}^0 d\zeta\, \sinh^2\zeta \phi'{}^2.
\end{equation}
Hence to the linear order, we find
\begin{equation}
  \label{Deleta}
  \Delta\eta
={\kappa\over2}\int_{-\infty}^0 d\zeta\, \sinh^2\zeta \phi'{}^2.
\end{equation}

Now let us solve for $\tbw^p$ to the first order in $\kappa{\phi'}^2$.
For this purpose, we set
\begin{equation}
  \tbw^p=e^{ip\eta}e^{F}\,;\quad 
F(\eta)=\int_{-\infty}^\eta d\eta'f(\eta')\,,
\label{bwp1}
\end{equation}
and insert this to Eq.~(\ref{tensoreq}). Linearizing the resulting
equation, we obtain
\begin{equation}
  \label{flineq}
  f'+ 2ipf={\kappa\over2}{\phi'}^2.
\end{equation}
This can be easily solved to give
\begin{equation}
  f={\kappa\over2}\int_{-\infty}^\eta d\eta'{\phi'}^2
e^{2ip(\eta'-\eta)} d\eta'.
\label{flinsol}
\end{equation}
Hence we obtain
\begin{eqnarray}
  \label{Flinsol}
F&=&{\kappa\over2}\int_{-\infty}^\eta d\eta'
\int_{-\infty}^{\eta'}d\eta''{\phi'}^2(\eta'')e^{2ip(\eta''-\eta')}
\nonumber\\
&=&{\kappa\over2}\int_{-\infty}^\eta{\sin p(\eta-\eta')\over p}
{\phi'}^2(\eta')e^{ip(\eta'-\eta)}d\eta',
\end{eqnarray}
where the second expression is obtained by integration by parts.

Then the spectrum of $Q_T/a^2$ can be evaluated by using
Eq.~(\ref{QTapprox}).  Our interest here is the behavior of $\tbw^p$
for small $p$. In particular, if the phase of $\tbw^p$ would remain
finite for $p\to0$, the tensor-type perturbation would give divergent
contribution to the CMB anisotropy as mentioned at the end of
subsection \ref{tenspec}. Hence it is important to clarify the
limiting behavior of $\tbw^p$ at $p\to0$.

In the limit $p\ll1$, Eq.~(\ref{Flinsol}) reduces to
\begin{equation}
  \label{Fapprox}
  F\approx{\kappa\over2}
\left[\int_{-\infty}^\zeta d\zeta'(\zeta-\zeta'){\phi'}^2(\zeta')
- ip
\int_{-\infty}^\zeta d\zeta'(\zeta-\zeta')^2{\phi'}^2(\zeta')\right],
\end{equation}
where we have replaced $\eta$ with $\zeta=\eta+\Delta\eta$.
The first term gives the first order correction to the amplitude of
$\tbw^p$. An inspection of it shows it diverges logarithmically at
$\zeta\to0$, which implies the behavior of $\tbw^p$ as
$\sim|\zeta|^{-\epsilon}$ where $\epsilon=O(\kappa\dot\phi^2/H^2)$.
 However, this divergence is precisely canceled by the correction 
to $H\approx h\approx Q$ which behaves as $\sim|\zeta|^\epsilon$.
Therefore the validity of Eq.~(\ref{traT}) remains intact.
On the other hand, the second term which gives a phase correction
which is finite at $\zeta\to0$. From Eqs.~(\ref{Deleta}) and (\ref{bwp1}),
we find
\begin{equation}
  \label{bwlimit}
  \tbw^p\to|\tbw^p|e^{ ip\beta(p)}\quad(\zeta\to0);
\quad \beta(p)=
{\kappa\over2}
\int_{-\infty}^0 d\zeta(\sinh^2\zeta-\zeta^2){\phi'}^2+O(p).
\end{equation}
Thus the phase $p\beta(p)$ vanishes for $p\to0$ as one should have
physically expected. 

Taking the above result into account, the final amplitude of
$Q^p_T/a^2$ is found to be
\begin{eqnarray}
\label{QTfinal}
\langle|U_{p\ell m}^{(+)}|^2\rangle &=&  |\tilde{\cal N}_p^T|^2
\left\langle\left|{Q_T^p\over a^2}\right|^2\right\rangle
=2|\tilde{\cal N}_p^T|^2\,{\kappa^2H^2|_{t=t_p}\over (p^2+1)^2}
\,\left[\cosh\pi p-|\rho_+|\cos\left(2p(\alpha-\beta)\right)\right]
\nonumber\\
&=&4 \kappa H^2|_{t=t_p}
{\cosh\pi p-|\rho_+|\cos\left(2p(\alpha-\beta)\right)
\over2 p(p^2+1)\sinh\pi p}\,,
\end{eqnarray}
where $\alpha$ and $\beta$ are functions of $p$, defined in
Eqs.~(\ref{rhoplus}) and (\ref{bwlimit}), respectively, and both are
finite in the limit $p\to0$. Thus the low frequency spectrum of 
the tensor-type perturbation is affected both by the configuration of
the Euclidean bubble and by the evolutionary behavior of the inflaton
field inside the bubble.

We note, however, that the effect of the evolution inside the bubble
 will be small when the small back reaction approximation is valid.
This comes from the fact that the integral in
 Eq.~(\ref{bwlimit}) can be basically evaluated over the first expansion
 times after nucleation, where $\sinh^2\xi$ dominates and can be replaced
 by $a$, hence $\beta\alt\kappa(\dot\phi^2)_{max}/H^2\ll 1$.
This implies the cosine starts oscillating only for $p \sim
 \beta^{-1} \gg 1$. Hence the effect of the evolution
 inside the bubble is unimportant if the small back reaction
 approximation is valid.
On the other hand, the effect may be significant for a model that
 violates the small back reaction approximation, and then the
general formula (\ref{generalT}) should be used. This typically
requires numerical solution for the reflection coefficient and
transfer functions, and we have no further comment for such a case here.

\section{conclusion}

We presented a formalism to calculate the scalar-type and 
tensor-type perturbations in the one-bubble open inflation scenario
and investigated the general features of their spectra. 
Our analysis is
based on a single-field model of one-bubble open inflation, but the
result for the tensor-type perturbation will be the same for a
multi-field model.

For the scalar-type perturbation, for a single-field model, we found
that there is no discrete supercurvature mode if the wall thickness is
sufficiently smaller than the wall radius. 
As for the continuous spectrum, we obtained the formula (\ref{scspec}),
which reduces to the familiar expression 
\begin{eqnarray*}
 \langle \vert{\cal R}^p_c\vert^2\rangle
\approx{9\over 2 p^3}\left.\left({H^3\over V'}\right)^2\right
   \vert_{t=t_p}, 
\end{eqnarray*}
for $p^2\gg 1$, where $t_p$ is approximately the horizon crossing time 
at which $p^2+1=H^2a^2$. In deriving it, we adopted the small back
reaction and assumed the small time variation of the potential curvature 
(i.e., the mass term). 
The small back reaction approximation is the
one in which the effect of the inflaton dynamics on the cosmological
expansion is small within a few Hubble expansion times and it
corresponds to the slow roll approximation at the late stage when the
spatial curvature of the universe can be neglected.
{}For $p^2\alt 1$, the resulting spectrum is 
largely dependent on details of a model. In particular, if the time
variation of the potential curvature is large, our formula
(\ref{scspec}) will not be a good approximation. 
However, the effect of the very low frequency modes ($p^2\ll1$) 
on the CMB anisotropy is known to be very small. 
Hence, we conclude that observational consequences of the 
scalar-type perturbations are rather robust except for the contribution
to the low multipoles ($\ell \alt$ a few) of the CMB anisotropy
spectrum, for which the $p^2\sim1$ part of the scalar
perturbation spectrum is important.

For the tensor-type perturbation, it was easy to see 
that there is no supercurvature mode. 
Again, the continuous spectrum is found to be 
almost independent of details of a model 
as long as the small back reaction condition is satisfied. 
By taking account of the back reaction perturbatively, 
we found the small $p$ behavior of the spectrum is affected
both by the configuration of the Euclidean bubble and by the evolution of
the inflaton field inside the bubble. For a model that satisfies the
small back reaction condition, however, the latter effect is found to be
unimportant. Nevertheless, by relaxing the small back reaction
condition, it may be of interest to see if this effect
gives rise to an additional constraint on models of one-bubble open
inflation.

\vspace{1cm} 
\centerline{\bf Acknowledgments} 

The work of M.S. and T.T. was supported in part by the Monbusho
Grant-in-Aid for Scientific Research No.~09640355 and by the Saneyoshi
foundation. J.G and X.M.  acknowledge support from CICYT under
contract AEN98. We acknowledge the use of CMBFAST for the computation
of the scalar and tensor temperature power spectra in Fig.~\ref{HTanis}.

\vspace{1cm}

\appendix

\section{canonical quantization of perturbations in one-bubble open
  inflation} 
Here, we discuss the quantization of perturbations of a scalar field 
and metric in the context of one-bubble open inflation. 
The Lorentzian continuation of the Coleman-De~Luccia instanton 
is considered as the background configuration. 
Following the Dirac's procedure, we reduce the degrees of freedom of 
a constrained system to physical ones. 
Here we follow the notation used in \cite{tamamilne}.
  %Paper I (OU-TAP 45, gr-qc/9610060). 

We start with the background metric of the form,
\begin{equation}
  ds^2=d\tau^2+a^2(\tau)(-d\chi^2+\cosh^2\chi d\Omega^2).
\end{equation}
For convenience, we introduce the unit normal vectors, 
\begin{equation}
 \tau^{\mu}:=(\partial_{\tau})^{\mu}=(1,0,0,0),\quad 
 n^{\mu}:=a^{-1}(\partial_{\chi})^{\mu}=(0,a^{-1},0,0). \quad 
\end{equation}
Then the metric is decomposed as 
\begin{equation}
 g_{\mu\nu}=\tau_{\mu}\tau_{\nu}-n_{\mu}n_{\nu}+\sigma_{\mu\nu}, 
\end{equation}
where $\sigma_{\mu\nu}$ is the metric of the 2-sphere with the 
radius $a$. Further we adopt the convention to denote the projection of
tensors as  
\begin{eqnarray}
 V_{\tau} & := & V_{\mu}\, \tau^{\mu},
\cr
 V_{n} & := & V_{\mu}\, n^{\mu}=a^{-1} V_{\chi}. 
\end{eqnarray}
The following relation is used in the calculations.
\begin{eqnarray}
 n^{\mu}{}_{;\nu}& = &-{\dot a\over a} \tau^{\mu}n_{\nu}
                   +{\tanh\chi \over a} \sigma^{\mu}{}_{\nu}, 
\cr
 \tau^{\mu}{}_{;\nu}& = &-{\dot a\over a} n^{\mu}n_{\nu}
                   +{\dot a \over a} \sigma^{\mu}{}_{\nu}.
\end{eqnarray}

The second variation of the Lagrangian 
for the gravitational part is given by  
\begin{equation}
L^{(2)}_G={1\over 8\kappa}\left(-h_{\mu\nu;\rho}h^{\mu\nu;\rho} 
  + 2 h_{\mu\nu;\rho}h^{\rho\mu;\nu}
  - 2 h_{\mu\nu}{}^{;\nu} h^{;\mu} + h_{;\mu} h^{;\mu}\right),
\label{A1}
\end{equation}
where $h_{\mu\nu}$ is the metric perturbation.
The matter part is given by 
\begin{eqnarray}
L^{(2)}_{mat} & = &
 {1\over 8\kappa} \left(h^2-2h_{\mu\nu} h^{\mu\nu}\right)
\left({\ddot a\over a}+2\left({\dot a\over a}\right)^2 -{2\over a^2}\right)
\cr
 && +{1\over 2}\left[\left(2h_{\tau}^{\nu}- h g_t^{\nu}\right) 
      \dot \phi \,\varphi_{,\nu}- h V'(\phi)\varphi 
      -g^{\mu\nu}\varphi_{,\mu}\varphi_{,\nu} 
      -V''(\phi)\varphi^2 \right], 
\end{eqnarray}
where $\varphi$ is the scalar field perturbation.
The explicit form of the terms in the gravitational part 
is given in Eq.~(E.5) of 
\cite{TStensor}. %Paper II (''The Spectrum of Gravitational Wave ...'')

We expand the metric and the scalar field perturbations in terms
of the spherical harmonics $Y=Y_{\ell m}(\Omega)$
and consider only the even parity modes.
For the metric components, we set
\begin{eqnarray}
&& h^{(e)}_{nn}=\sum H^{(e)\ell m}_{nn}Y, \quad 
   h^{(e)}_{n\tau}=\sum H^{(e)\ell m}_{n\tau}Y, \quad 
   h^{(e)}_{\tau\tau}=\sum H^{(e)\ell m}_{\tau\tau}Y, 
\cr
&& h^{(e)}_{nA}=\sum H^{(e)\ell m}_{n}Y_{||A},\quad 
   h^{(e)}_{\tau A}=\sum H^{(e)\ell m}_{\tau}Y_{||A},
\cr 
&& h^{(e)}_{AB}=\sum \left(w^{(e)\ell m} Y \hat\sigma_{AB}
                 +v^{(e)\ell m} Y_{AB}\right),
\label{vardef}
\end{eqnarray}
where 
\begin{equation}
 Y_{AB}={Y_{||AB}\over \ell(\ell+1)}+{1\over 2}\hat\sigma_{AB} Y. 
\end{equation}
The reality condition implies 
$\overline{H_i^{\ell m}}=H_i^{\ell -m}$, where
$H_i=H_{nn}$, $H_{n\tau}$, $H_{n}$, $H_{\tau\tau}$, $H_{\tau}$, $w$, $v$. 
To keep the simplicity of notation, we omit the indices,
$(e)$, $\ell$ and $m$, unless there arises confusion. 
For later convenience, 
we list the formulas of the $\Omega$-integration, 
\begin{eqnarray}
 \int d\Omega~Y\overline{Y}&=&1,
\cr
 \int d\Omega~\hat\sigma^{AA'}Y_{||A}\overline{Y_{||A'}}
 &=&{\ell(\ell+1)},
\cr
 \int d\Omega~\hat\sigma^{AA'}\hat\sigma^{BB'}Y_{AB}\overline{Y_{A'B'}}&=&
                   {\ell(\ell+1)-2\over 2\ell(\ell+1)},
\cr
 \int d\Omega~\hat\sigma^{AA'}\hat\sigma^{BB'}\hat\sigma^{CC'}
   Y_{AB||C}\overline{Y_{A'B'||C'}}
      &=&{(\ell(\ell+1)-2)(\ell(\ell+1)-4)\over 2\ell(\ell+1)},
\cr
 \int d\Omega~\hat\sigma^{AA'}\hat\sigma^{BB'}\hat\sigma^{CC'}
   Y_{AB||C}\overline{Y_{A'C'||B'}}&=&
                   {(\ell(\ell+1)-2)(\ell(\ell+1)-6)\over 4\ell(\ell+1)}. 
\label{omegaint}
\end{eqnarray}
It is convenient to rewrite the components having more 
than two of their indices projected onto the 2-sphere as
\begin{eqnarray}
  h_{nA;B}
&=& \left[-{\ell(\ell+1)\over 2 a^2\cosh^2\chi} H_n
         +{1\over a}H_{n\tau}-{\tanh\chi\over a} 
           \left(H_{nn}+{w\over a^2\cosh^2\chi}\right)
         \right]\sigma_{AB} Y
\cr && \quad
   +\left[\ell(\ell+1)H_n -{\tanh\chi\over a}~v\right] Y_{AB},
\cr
  h_{\tau A;B}
&=& \left[-{\ell(\ell+1)\over 2 a^2\cosh^2\chi} H_{\tau}
         +{1\over a}\left(H_{\tau\tau}-{w\over a^2\cosh^2\chi}\right)
          -{\tanh\chi\over a} H_{n\tau}\right]\sigma_{AB} Y
\cr && \quad
   +\left[\ell(\ell+1)H_{\tau} -{1\over a}~v\right] Y_{AB},
\cr
 h_{AB;n}
&=& {\sigma_{AB} Y \over a^3\cosh^2\chi}
   \left[\partial_{\chi}-{2\tanh\chi}\right] w
         +{1\over a}Y_{AB} \left[\partial_{\chi}-{2\tanh\chi}\right] v,
\cr
 h_{AB;\tau}
&=& {\sigma_{AB} Y \over a^2\cosh^2\chi}
   \left[\partial_{\tau}-2{\dot a\over a}\right] w
         +Y_{AB} \left[\partial_{\tau}-2{\dot a\over a}\right] v,
\cr
 h_{AB;C}
&=&
 {w\over a^2\cosh^2\chi} Y_{||C}\sigma_{AB}
 +v~ Y_{AB||C}
 +{2\over a}\left[H_{\tau}-\tanh\chi H_n\right]
  Y_{||(A}\sigma_{B)C}.
\end{eqnarray}

Then it is straightforward to calculate the action for the even parity
modes. 
By using the formulas (\ref{omegaint}) the $\Omega$-integration in the
action is performed to give 
\begin{equation} 
 \int d^4 x\,\sqrt{-g}\,\left(L^{(2)}_{G}+L^{(2)}_{mat}\right)
=\int d\chi \int d\tau\,\left({\bf L}^{(e)}_{G}+{\bf L}^{(e)_{mat}}\right).
\end{equation} 
Here we only demonstrate 
the most complicated term in the gravitational part; 
\begin{eqnarray}
&& \int d\Omega~
       \sigma^{\mu\mu'} \sigma^{\nu\nu'}\sigma^{\rho\rho'} 
       \left(-h_{\mu\nu;\rho}h_{\mu'\nu';\rho'}
       +2h_{\mu\nu;\rho}h_{\nu'\rho';\mu'}\right)
\cr
&&\quad \quad
 =\sum_{\ell,m}{1\over (a^2\cosh^2\chi)^3}
  \Biggl[-{\ell(\ell+1)-2\over\ell(\ell+1)} \vert v\vert^2
  +4\ell(\ell+1)(a\cosh^2\chi)^2
    \left\vert H_{\tau}-\tanh\chi H_n\right\vert^2
\cr
&&\quad\quad\quad\quad
  +8\ell(\ell+1)a \cosh^2\chi \left[H_{\tau}-\tanh\chi H_n\right]
 \overline{w}
  -2 \left(\ell(\ell+1)-2\right) v~\overline{w}\Biggr].
\end{eqnarray}
The matter part reduces to 
\begin{eqnarray}
 {\bf L}^{(e)}_{mat}={1\over 4\kappa a^3\cosh^2\chi}
    && \Biggl[- \left( H_{nn}^2 + H_{\tau\tau}^2 + 
       {1\over a^4\cosh^4 \chi }
           \left( 2w^2+ 
             {l(l+1)-2\over 2l(l+1)}v\right)\right)
\cr &&
    \quad  +2\left( H_{n\tau}^2 + 
       {l(l+1) \over a^2 \cosh^2 \chi}
        \left(H_n^2 - H_{\tau}^2 \right) \right)  
       +{1\over 2} (\hbox{\rm Tr} H)^2  \Biggr]
      \left( {\ddot a\over a} + 
    2\left({\dot a\over a}\right)^2 - {2\over a^2}\right) 
\cr &&
  +4\Biggl[-2\left[-H_{n\tau} {1\over a}\partial_{\chi} X + 
     \left(H_{\tau\tau}- {\hbox{\rm Tr} H\over 2} \right)
     \left(\dot X + fX\right)
     +{l(l+1)\over a^2\cosh^2\chi} H_{\tau}X\right] 
\cr && 
   \quad +\left(f+3{\dot a\over a}\right)(\hbox{\rm Tr} H)X
    +(\dot X+f X)^2-{1\over a^2} \left(\partial_{\chi}X\right)^2 
\cr&&
   \quad+\left({l(l+1)\over a^2\cosh^2\chi}+\dot f 
     +3{\partial({\dot a/ a})\over \partial\tau}
     +f\left(f+3{\dot a\over a}\right)\right) X^2\Biggr]
     \left({\ddot a\over a}-\left({\dot a\over a}\right)^2
     +{1\over a^2}\right),
\end{eqnarray}
where 
\begin{equation}
 \hbox{\rm Tr} H= -H_{nn} +H_{\tau\tau} +{2 w\over a^2\cosh^2\chi}, 
\end{equation}
\begin{equation}
 X:={\varphi\over \dot \phi}, 
\end{equation}
and 
\begin{equation}
   f(\tau):={\ddot \phi\over \dot \phi}.
\end{equation}
Here we note that there is a relation: 
\begin{equation}
 b:={\ddot a\over a}-\left({\dot a \over a}\right)^2 +{1\over a^2}
 =-{\kappa\over 2}\dot\phi^2,
\end{equation}
which is used to remove $\dot\phi^2$ from the action. 
As a result, the action contains $\kappa$ as an overall factor. 
We introduced $f$ just for notational simplicity. 
In fact, it can be written in terms of $b$ (or  $a$), as 
\begin{equation}
 f={1\over 2}{d\over d\tau}\log b.
\end{equation}
Hence the action is rewritten into the form that depends 
on the background quantity only through the scale factor $a$. 
This fact simplifies the calculation considerably. 
In addition, for computational simplicity, we set $4\kappa=1$ in the
rest of this appendix.

Next we define the canonical conjugate momenta by 
\begin{equation}
 \overline{P^{(e)\ell m}_i} :={\partial{\bf L}^{(e)}\over\partial
(\partial_\chi H^{(e)\ell m}_i)},
\end{equation}
where 
$P_i=P_{nn}$, $P_{n\tau}$, $P_{n}$, $P_{\tau\tau}$, $P_{\tau}$, $P_w$,
$P_v$, $P_{X}$. 
Since the $\chi$-derivatives of $H_{nn}$, $H_{n\tau}$ and $H_{n}$ 
are not contained in the defining equations of the conjugate momenta, 
they give the primary constraint equations: 
\begin{eqnarray}
 C_1 & := & P_{nn}-l(l+1) H_{n}-2a \cosh\chi\sinh\chi H_{nn}
     -{2\tanh\chi\over a} w
     +a^2\cosh^2\chi\left[\partial_{\tau}+2{\dot a\over a}\right]
 H_{n\tau}=0,
\cr
  C_2 & := & P_{n\tau}-a^2\cosh^2\chi \partial_\tau (H_{nn}+H_{\tau\tau})
    -2\left[\partial_{\tau}-2{\dot a\over a}\right]w =0,
\cr
  C_3 & := & P_{n}-\ell(\ell+1)\left(H_{nn}+H_{\tau\tau}
      +{2\over a^2\cosh^2\chi}w-{4\tanh\chi\over a} H_{n}\right)=0.
\end{eqnarray}
The other components are
\begin{eqnarray}
  P_{\tau\tau} &=& -{2\over a}\partial_\chi w
  -\ell(\ell+1) H_{n}-2a\cosh\chi\sinh\chi H_{nn}+{2\tanh\chi\over a} w 
  -a^2\cosh^2\chi \left[\partial_{\tau}-2{\dot a\over a}\right]H_{n\tau},
\cr
  P_{\tau} &=& 2\ell(\ell+1) 
  \left({1\over a} \partial_\chi H_{\tau}-H_{n\tau}-
    \left[\partial_{\tau}-{\dot a\over a}\right] H_{n}\right),
\cr
  P_{w} &=& -{2\over a^3\cosh^2\chi}\left[\partial_\chi-2\tanh\chi \right] w
            -{2\over a} \partial_\chi H_{\tau\tau} 
            +2\left[\partial_{\tau}+2{\dot a\over a}\right]H_{n\tau},
\cr
  P_{v} &=& {\ell(\ell+1)-2\over 2\ell(\ell+1)}{1\over a^3\cosh^2\chi} 
             \partial_\chi v
            -{\ell(\ell+1)-2\over a^2\cosh^2\chi} H_{n}, 
\cr 
  P_{X} & =& -8 b \dot \phi^2 a\cosh^2\chi 
     \left(\partial_{\chi} {X}- a H_{n\tau}\right). 
\end{eqnarray}
The Hamiltonian is defined by 
\begin{equation}
 {\bbox h}^{(e)}=\sum_{\ell, m}\sum_i \overline{P^{(e)}_{i}} 
    \left(\partial_\chi H^{(e)}_{i}\right)-{\bf L}^{(e)},  
\end{equation}
where $\partial_\chi H_{nn}$, $\partial_\chi H_{n\tau}$ and 
$\partial_\chi H_{n}$ are to be replaced 
by $\lambda_1, \lambda_2$ and $\lambda_3$, respectively. 
The canonical equations of motion are 
\begin{eqnarray}
 H_{i} & = & {\partial {\bbox h}\over \partial \overline{P_{i}}}, 
\cr
 {P_{i}} & = & -{\partial {\bbox h}\over \partial \overline{H_{i}}}.
\label{a17}
\end{eqnarray}

The consistency conditions for primary constraints are calculated as 
\begin{eqnarray}
 \partial_{\chi} C_1 & = & 2a\left[\partial_{\tau}^2
    -{\dot a\over a}\partial_{\tau}
    +\left(2{\ddot a\over a}-{l(l+1)-2\over 2a^2\cosh^2\chi}\right)\right]w
    -4 b a^3\cosh^2\chi \left[\partial_{\tau}
    +2f+3{\dot a\over a}\right] {X} 
\cr
 && -2a^3\cosh^2\chi \left[{\dot a\over a}\partial_{\tau} 
    +{\ddot a\over a}+2\left({\dot a\over a}\right)^2 
    -{1\over a^2} +{l(l+1)\over 2 a^2\cosh^2\chi}\right] H_{\tau\tau}
 +2a^2 \cosh\chi\sinh\chi \left[\partial_{\tau}
   +2{\dot a\over a} \right] H_{n\tau}
\cr
 && +2l(l+1) a \left[ \partial_{\tau} 
    +{\dot a\over a} \right] H_{\tau}
    -2a\sinh^2\chi H_{nn} -2l(l+1) \tanh\chi H_n 
\cr 
 && -{(l(l+1)-2)v\over 2a \cosh^2\chi}
   +a^2\cosh\chi\sinh\chi P_w=0, 
\end{eqnarray}
\begin{eqnarray}
 \partial_{\chi} C_2 & = & 2a\partial_{\tau} P_{\tau\tau}
    +2a^3\cosh^2\chi\left[\partial_{\tau}^2
    +2{\dot a\over a}\partial_{\tau}
    +2{\ddot a\over a}-2\left({\dot a\over a}\right)^2\right]H_{n\tau}
    -2l(l+1) a\left[\partial_{\tau}
    -2{\dot a\over a}\right] H_n 
\cr
 && -2a^2\dot a\cosh^2\chi P_w 
     -a (P_{\tau}+P_{X})+4 a\dot a \cosh\chi\sinh\chi H_{nn} 
    + 4{\dot a\over a}\tanh\chi \, w=0, 
\end{eqnarray}
and
\begin{eqnarray}
 \partial_{\chi} C_3 & = & 
    a \left[\partial_{\tau} 
    +2{\dot a\over a}\right]\left(P_{\tau}+2l(l+1) H_{n\tau}\right) 
    -4l(l+1)\tanh\chi \left[\partial_{\tau}  
   +{\dot a\over a}\right] H_{\tau}
\cr && 
   + l(l+1)\left[2\tanh\chi(H_{\tau\tau}-H_{nn})
     -2{l(l+1)\over a\cosh^2\chi} H_n +a(P_w-2 P_v)\right]
     +2(l(l+1)-2){\sinh\chi\over a^2\cosh^3\chi} v=0.  
\end{eqnarray}
These can be solved for $H_{nn}$, $P_v$ and $P_{X}$.  
Further consistency conditions, $\partial_\chi^2 C_i=0$, 
become trivial. 

We have to mention that 
all the above equalities hold in the weak sense. 
That is, in reducing the expression 
we used the primary constraints and their consistency conditions 
that have been already obtained. 
Since the set of primary constraints and 
their consistency conditions closes, 
we can substitute them into the action, and 
remove the six variables, $P_{nn}$, $P_{n\tau}$, $P_{\tau\tau}$, 
$H_{nn}$, $P_v$ and $P_{X}$. 

There still remain unphysical gauge degrees of freedom. 
In order to obtain the reduced action that contains only the 
physical degrees of freedom, we set the following 
gauge condition corresponding to the Newton gauge: 
\begin{equation}
 2X=-b^{-1}\left(\partial_{\tau}+{\dot a\over a}\right) 
    H_{\tau\tau},\quad H_{n\tau}=0,\quad H_{\tau}=0. 
\label{Gcondi}
\end{equation}
These gauge conditions imply the consistency conditions 
$
\quad \partial_\chi \left[
2X+b^{-1}\left(\partial_{\tau}+{\dot a \over a}\right)
H_{\tau\tau}\right]=0,
\quad \partial_\chi H_{n\tau}=0$ and  
$\partial_\chi H_{\tau}=0$, 
which respectively become 
\begin{equation}
 \partial_{\tau} D_1 = 0, \quad 
 \lambda_2=0, \quad
 D_3=0,
\end{equation}
where
\begin{eqnarray}
 D_1 & = & 
    \left[\partial_{\tau}^2 -{\dot a\over a} \partial_{\tau}
      -2{\ddot a\over a}+{1\over a^2} 
     -{l(l+1)\over 2a^2 \cosh^2\chi}\right] w
    +a^2\cosh^2\chi \left[\partial_{\tau}^2 
      +3{\dot a\over a} \partial_{\tau}
      +{1\over a^2} -{l(l+1)\over 2a^2 \cosh^2\chi}\right] H_{\tau\tau}
 \cr &&    
     -{l(l+1)\tanh \chi\over 2a} H_n 
     -{(l(l+1)-2)  \over 4a^2\cosh^2\chi} v,   
\cr
 D_3 & = & {a P_{\tau}\over 2l(l+1)} +a\left[
    \partial_{\tau}-{\dot a\over a} \right] H_n. 
\end{eqnarray}
{}From the condition $\partial_{\tau} D_1=0$, we can set 
$D_1$ as an arbitrary function of $\chi$. This is due 
to the fact that the gauge condition (\ref{Gcondi}) does 
not completely fix the gauge. Therefore we must impose 
an additional condition. Here we choose $D_1=0$. 
The first two equations $D_1=0$ and $D_3=0$ can be solved 
for $P_{\tau}$ and $H_{n}$. 

As the expression for $D_1$ is rather complicated, 
we first consider the consistency condition for $D_3$. 
The condition $\partial_{\chi} D_3=0$ determines $\lambda_3$ as 
\begin{eqnarray}
  \lambda_3 & = & 
    {a\over \sinh^2\chi} 
    \left[\partial_{\tau}^2 -{\dot a\over a} \partial_{\tau}
      -2{\ddot a\over a}-{1\over a^2} 
       -{l(l+1)-4\over 2a^2 \cosh^2\chi}\right] w
    +a^3\coth^2\chi\left[\partial_{\tau}^2 +3{\dot a\over a}\partial_{\tau}
      -{1\over a^2} -{l(l+1)-4\over 2a^2 \cosh^2\chi}\right] H_{\tau\tau}
 \cr &&    
     -2\tanh\chi\left(1+{l(l+1)\over 2\sinh^2\chi}\right) H_n 
     -{(l(l+1)-2)  \over 2l(l+1) a\cosh^2\chi}
      \left(1+{l(l+1)\over 2\sinh^2\chi}\right) v 
     +{a^2\coth \chi\over 2} P_w +\alpha(\chi) a,  
\end{eqnarray}
where $\alpha(\chi)$ is an arbitrary function of $\chi$ which 
 arises due to the remaining gauge degrees of freedom. 
Here we simply choose $\alpha(\chi)=0$. 

Using the relations which have been already obtained, the 
second level consistency conditions for the first gauge condition, 
$\partial_\chi D_1=0$ reduces to 
\begin{equation}
\left[\partial_{\tau}^2 +5{\dot a\over a} \partial_{\tau}
      +{\ddot a\over a}+3\left({\dot a\over a}\right)^2
     +{1\over a^2} 
   -{l(l+1)\over 2a^2 \cosh^2\chi}\right]E_1=0,
\end{equation}
where
\begin{eqnarray}
 E_1 & := &P_{w} - {2\over a\sinh\chi\cosh\chi} 
    \left[\partial_{\tau}^2 -{\dot a\over a} \partial_{\tau}
      -2{\ddot a\over a}+{4\over a^2} 
         -{l(l+1)+6\over 2a^2 \cosh^2\chi}\right] w
 \cr &&    
    -2a\coth\chi \left[\partial_{\tau}^2 +3{\dot a\over a} \partial_{\tau}
      +{4\over a^2} -{l(l+1)+6\over 2a^2 \cosh^2\chi}\right] H_{\tau\tau}
     +{(l(l+1)-2)  \over 2a^3\sinh\chi\cosh^3\chi} v. 
\end{eqnarray}
Again by choosing the simplest choice $E_1=0$, 
we obtain the equation which determines $P_w$. 
Furthermore $\partial_\chi E_{1}=0$ gives the condition,
\begin{eqnarray}
 \lambda_1 &=&-{2\over \sinh\chi\cosh\chi} 
    \left[\partial_{\tau}^2 -{\dot a\over a} \partial_{\tau}
      -2{\ddot a\over a}+{4\over a^2} 
     -{l(l+1)+6\over 2a^2 \cosh^2\chi}\right] w
\cr&&
    -2a^2\coth\chi \left[\partial_{\tau}^2 +3{\dot a\over a} \partial_{\tau}
      +{4\over a^2} -{l(l+1)+6\over 2a^2 \cosh^2\chi}\right] H_{\tau\tau}
     +{(l(l+1)-2)  \over 2a^2\sinh\chi\cosh^3\chi} v 
     +{1\over 2 a\cosh^2\chi} P_{\tau\tau}. 
\end{eqnarray}

Now we find that the set of all the constraints closes and 
it becomes second class. Hence we can remove the 
unphysical variables by using these constraints. 
Then the remaining variables are 
$P_{\tau\tau}$, $w$, $v$ and $H_{\tau\tau}$.  
We define the scalar-type and tensor-type variables 
by 
\begin{eqnarray}
 Q_S^{(e)\ell m} & := &4a H^{(e)\ell m}_{\tau\tau},\cr
 Q_T^{(e)\ell m} & := & {a^2 \cosh^2\chi\over 2} 
     \left[H^{(e)\ell m}_{nn}\right]^{(\hbox{TF})}
        = a^2 \cosh^2\chi H^{(e)\ell m}_{\tau\tau} +w^{(e)\ell m}, 
\end{eqnarray}
where $(\hbox{TF})$ means the trace free part and 
we have used the fact that $H_{nn}$ is rewritten by using 
the set of second class constraints as
\begin{equation}
 H_{nn}=3 H_{\tau\tau} +{2 w\over a^2\cosh^2\chi}. 
\end{equation}

Then the equations for $Q_S$ and $Q_T$ reduce to 
\begin{eqnarray}
 \partial_\chi Q_S & = & {\Pi_S\over \cosh^2\chi},
\cr
 \partial_\chi \Pi_S & = & \left[-\ell(\ell+1)
   +(3-\widehat K_S)\cosh^2 \chi\right] Q_S, 
\cr 
 \partial_\chi Q_T & = & {\Pi_T\over \cosh^2\chi},
\cr
 \partial_\chi \Pi_T & = & \left[-\ell(\ell+1)
   -\widehat K_T\cosh^2 \chi\right] Q_T, 
\label{a30}
\end{eqnarray}
where $\Pi_S$ and $\Pi_T$ are defined by 
\begin{eqnarray}
 \Pi_S^{(e)\ell m}
 & := & 2P_{\tau\tau}^{(e)\ell m},
\cr
 \Pi_T^{(e)\ell m}
 & := & -\cosh^2 \chi\left[{\ell(\ell+1)a}H^{(e)\ell m}_{n}
   +2 \tanh\chi~ (w^{(e)\ell m}+a^2\cosh^2\chi 
       H_{\tau\tau}^{(e)\ell m}\right], 
\end{eqnarray}
and the derivative operators $\widehat K_S$ and $\widehat K_T$ are
defined by  
\begin{eqnarray}
 \widehat K_S &:= & \left[-a^3\dot\phi^2{\partial\over\partial\tau}
   {1\over a\dot\phi^2}{\partial\over\partial\tau}  
     -a^2 b\right], 
\cr
  \widehat K_T &:= & \left[-a{\partial\over\partial\tau}
   a^3{\partial\over\partial\tau}{1\over a^2}\right]. 
\end{eqnarray}

All the other variables can be written in terms of $Q_{T/S}$ and
$\Pi_{T/S}$ as
\begin{eqnarray}
 H_{\tau\tau}& = &{Q_S\over 4a},
\cr
 H_{nn} & = & {Q_S\over 4a}+{2Q_T\over a^2\cosh^2\chi},
\cr
 H_{n} & = & -{2\over \ell(\ell+1)a}\left[
             \tanh\chi~ Q_T+{\Pi_T\over \cosh^2\chi}\right],
\cr
 H_{\tau} & = & H_{n\tau}=0,
\cr
 v & = & {4\cosh^2\chi\over \ell(\ell+1)-2}\left[
        \left(2-\widehat K_T
-{\ell(\ell+1)+2\over 2\cosh^2\chi}\right) Q_T
         +{\tanh\chi\over \cosh^2\chi} \Pi_T\right],
\cr
 w & = & Q_T + {a\cosh^2\chi\over 4} Q_S, 
\cr
 {X} & = & -{{\partial\over\partial\tau} Q_S\over 8ab}
\cr
 P_{nn} & = & {2\over a}\left[2\tanh\chi~Q_T-{\Pi_T\over\cosh^2\chi}\right],
\cr
 P_{n} & = & {8\over a^2}\left[
        \left(1+{\ell(\ell+1)-2\over 2\cosh^2\chi}\right) Q_T
         +{\tanh\chi\over \cosh^2\chi} \Pi_T\right],
\cr
 P_{n\tau} & = & 4\left({\partial\over\partial\tau}
-2{\dot a\over  a}\right) Q_T, 
\cr
 P_{\tau\tau} & = & {\Pi_S\over 2},
\cr
 P_{\tau} & = &  {4\over a}
     \left({\partial\over\partial\tau}-2{\dot a\over a}\right) 
     \left(\tanh\chi~ Q_T+{\Pi_T\over\cosh^2\chi}\right),
\cr
 P_{w} & = &  {2\over a^3\cosh^2\chi} 
     \left(2\tanh\chi~ Q_T -{\Pi_T\over\cosh^2\chi}\right),
\cr
 P_{v} & = &  {2\over \ell(\ell+1)a ^3}
     \left[\left(4-3\widehat K_T 
-{2\over \cosh^2 \chi}\right)\tanh\chi~ Q_T 
     +\left(2-\widehat K_T +{\ell(\ell+1)-4\over 2\cosh^2 \chi}\right)
      {\Pi_T\over\cosh^2\chi}\right],
\cr 
 P_{X} & = & {\partial\over\partial\tau} \Pi_S.
\label{a31}
\end{eqnarray}
Of course, Eqs.~(\ref{a30}) and (\ref{a31}) 
are consistent with the equations that the mode functions satisfy.

Substituting (\ref{a31}) into the canonical form of the action 
\begin{eqnarray}
 \int d\chi \int d\tau~{\cal L}^{(e)\ell m}:=\int d\chi \int d\tau~ 
 \left(\sum_{\ell,m}\sum_i \overline{P_i^{(e)\ell m}} 
  \left(\partial_\chi H^{(e)\ell m}_i\right) -{\bbox h}^{(e)}\right),  
\label{canL}
\end{eqnarray}
 the reduced action becomes
%\footnote{Here we do not have to calculate 
%all terms in ${\cal L}$ because 
%we know the equation of motion which follows from the reduced action. 
%Therefore we have only to calculate the first term in the right 
%hand side of Eq.~(\ref{canL}). }
\begin{eqnarray}
 \int d\chi \int d\tau~{\cal L}^{(e)}_{(red)}
  = && \sum_{\ell,m}{8\over (\ell-1)\ell(\ell+1)(\ell+2)}
    \int d\chi \int {d\tau\over a^3} 
 \Biggl[
    \overline{\Pi_T}
 \widehat K_T(\widehat K_T -1)\left(\partial_\chi Q_T\right)
\cr &&\quad\quad
  -{1\over 2}\left({1\over\cosh^2\chi}\overline{\Pi_T}
   \widehat K_T(\widehat K_T -1)\Pi_T+
    \overline{Q_T} \widehat K_T(\widehat K_T -1)
    \left\{\ell(\ell+1)+\widehat K_T\cosh^2\chi
    \right\} Q_T \right)\Biggr]
\cr 
 && + \sum_{\ell,m}
    \int d\chi \int {d\tau \over a^3\dot\phi^2} 
\cr
 && \times 
\left[
    \overline{\Pi_S} \widehat K_S \left(\partial_\chi Q_S\right)-
  {1\over 2}\left({1\over\cosh^2\chi}\overline{\Pi_S}
   \widehat K_S \Pi_T+
    \overline{Q_S}\widehat K_S \left\{\ell(\ell+1) 
+(\widehat K_S -3)\cosh^2\chi
        \right\} Q_S\right)\right].
\end{eqnarray}

\section{Some bounds on W in the thin wall case}

For the case of thin wall bubbles, the scalar field 
ceases to move in the true and false vacuum limits. 
Then the geometry can be approximated by a 
pure de Sitter one. 
Comparing Eqs.~(\ref{ewq2}) and (\ref{pop}), we have 
\begin{equation}
 W'+W^2=4+{\mu^2\over \cosh^2\eta},
\label{Wprime}
\end{equation}
where
\begin{eqnarray*}
\mu^2={m^2\over H^2}-2,
\end{eqnarray*}
and
$m^2=d^2V/d\phi^2(\eta=\pm\infty)$.

Solving Eq. (\ref{Wprime}) perturbatively, the asymptotic behavior
of $W$ is found as
\begin{eqnarray}
  \label{asympW}
  W\to\pm2\pm{2\over3}\mu^2e^{\pm2\eta}\quad{\rm for}
\quad\eta\to\mp\infty.
\label{Wasymp}
\end{eqnarray}
{}From Eq.~(\ref{Wprime}), we can prove that 
\begin{equation}
W>2
\label{WW3}
\end{equation}
is maintained 
in the false vacuum side under the condition, $\mu^2>0$. 
The proof is as follows. 
If $W$ could become equal to or smaller than $2$, there would be a point
$\eta=\eta_c$ at which $W=2$ for the first time and $W'\leq0$ there.
But this contradicts with Eq.~(\ref{Wprime}). 
In the true vacuum side, the mass is not necessarily
large compared with $H$. As a minimum requirement, we shall
assume that $m^2>0$ in the true vacuum 
side. If $m^2=0$, we can solve (\ref{prime1}) as 
\begin{eqnarray*}
 \phi'&\propto& 2\cosh^2\eta -3\cosh\eta\sinh\eta+\sinh^2\eta\tanh\eta
  ={3+e^{-2\eta}\over1+e^{2\eta}}
\\
   &\propto&{\cosh(\eta+\delta)\over\cosh\eta}e^{-2\eta}\,,
\end{eqnarray*}
where a fixed number $\delta:=\ln\sqrt{3}>0$ is introduced.
Then $W_0=W|_{m^2=0}$ is given by
\begin{eqnarray}
  W_0=\tanh(\eta+\delta)-\tanh\eta-2\,.
\label{W0}
\end{eqnarray}
Now, let us show $W<W_0$ when $\mu^2>-2$. 
Namely, we prove that in true vacuum 
\begin{eqnarray}
  g=W-W_0<0.
\label{WTbound}
\end{eqnarray}
In fact, from Eq.~(\ref{asympW}) we have
\begin{eqnarray*}
  g\to-{2\over3}(\mu^2+2)e^{-2\eta}<0\quad{\rm for}\quad\eta\to\infty.
\end{eqnarray*}
Now if $g=0$ at $\eta=\eta_c$ as we decrease $\eta$ from infinity, we
must have $g'\leq0$ there. But
\begin{eqnarray}
  g'|_{g=0}={\mu^2+2\over\cosh^2\eta_c}>0,
\end{eqnarray}
which is a contradiction. Thus $g<0$ in true vacuum.
Since $W_0<0$, this also implies 
\begin{eqnarray}
  W^2>W_0^2.  
\label{WW0}
\end{eqnarray}
Let us denote the wall region as the region where neither 
(\ref{WW3}) nor (\ref{WW0}) is satisfied. Then it is straightforward
to show that, provided that the wall is located at $\eta>0$, the integral
(\ref{swkbcond}) is always smaller than $\pi$ provided that the wall
region has a width $\Delta\eta<<1$.

\section{An exactly soluble model}

Here we present a model in which both the scalar- and tensor-type
perturbation equations for the evolution inside the bubble
can be solved analytically. We follow the notation used in sections IV
and V.

Consider the case,
\begin{eqnarray}
  \phi'={\mu\over2\sinh2\zeta}\,.
\label{phip2}
\end{eqnarray}
In this case, we can solve Eq.~(\ref{Qeq}) for $Q$ as
\begin{eqnarray*}
  Q=Q_0(-\tanh\zeta)^\epsilon\,,
\end{eqnarray*}
where
\begin{eqnarray*}
  \epsilon={-1+\sqrt{1+4A^2}\over2}\,,\quad A^2:={\kappa\over32}\mu^2\,.
\end{eqnarray*}
{}From Eq.~(\ref{htoQ}), we have 
\begin{eqnarray*}
  h_0^2:=h^2(\zeta\to -\infty)=Q^2_0(1+2\epsilon).
\end{eqnarray*}
Therefore
\begin{eqnarray}
  Q={h_0\over\sqrt{1+2\epsilon}}(-\tanh\zeta)^\epsilon.
\label{Qsol}
\end{eqnarray}

In the limit $\zeta\to-\infty$
\begin{eqnarray*}
 a\to{2\over Q_0} e^{\zeta}={2\over h_0} e^{\eta}\,,
\end{eqnarray*}
where in the last equality we assume that the scale factor takes the de
Sitter form. Thus we find $\Delta\eta := \eta-\zeta$ as
\begin{eqnarray*}
 e^{2\Delta\eta}= 1+2\epsilon. 
\end{eqnarray*}

It is worthwhile to analyze the general behaviors of $\phi$ and $a$
near the nucleation point, i.e., in the limit $\eta\to-\infty$.
If we expand 
\begin{eqnarray}
 \phi & = & \phi_0+\phi_2 e^{2\eta}+\phi_4 e^{4\eta}+\cdots, \cr
 a & = & {2\over h_0} e^{\eta}
 \left(1 +a_2 e^{2\eta}+\cdots\right), 
\label{expand}
\end{eqnarray}
we obtain
\begin{eqnarray}
 a_2 & = & 1, \cr
 \phi_2 & = & -{3\over 2\kappa}{\partial V_0\over V_0}, \cr
 \phi_4 & = & {\phi_2\over 6}\left(2-{\partial^2V_0\over h_0^2}\right), 
\label{coef}
\end{eqnarray}
where the suffix $0$ denotes a quantity at $\eta\to-\infty$.

The scalar field potential can be reconstructed as follows.
Integrating Eq.~(\ref{phip2}), we have
\begin{eqnarray*}
  \phi={\mu\over4}\ln|\tanh\zeta|,,
\end{eqnarray*}
where we have set $\phi=0$ at the nucleation point.
Thus
\begin{eqnarray*}
  -\tanh\zeta=e^{4\phi/\mu}.
\end{eqnarray*}
For definiteness, we assume $\mu<0$, which implies $\phi$ rolls toward
its positive direction. Let us first examine the meaning of the
parameter $\epsilon$. In terms of $\epsilon$, $\mu$ is expressed as
\begin{eqnarray*}
  \mu=-4\sqrt{2\epsilon(1+\epsilon)\over\kappa}\,.
\end{eqnarray*}
We note that, from Eqs.~(\ref{expand}) and (\ref{coef}), we have
\begin{eqnarray*}
  \mu={3\partial V_0\over\kappa V_0}(1+2\epsilon)\,,
\end{eqnarray*}
which implies
\begin{eqnarray*}
  {\partial V_0\over V_0}
=-\sqrt{{32\kappa\epsilon(1+\epsilon)\over9(1+2\epsilon)^2}} \,,
\end{eqnarray*}
where $V_0$ and $V'_0$ are the values of the potential
and its $\phi$-derivative, respectively, at the nucleation point.
Furthermore, expanding $\phi$ to the next order in $e^{2\zeta}$,
we find $\phi_4=0$ in the present case. Hence
\begin{eqnarray*}
  M^2_0:=\partial^2V_0=2h_0^2\,.
\end{eqnarray*}
Thus the curvature of the potential is relatively large at the
nucleation point.

Now we reconstruct the potential.
{}From Eqs.~(\ref{defh}) and (\ref{htoQ}) together with
Eq.~(\ref{Qsol}), the potential $V$ is given by
\begin{eqnarray}
  V&=&{3\over\kappa}Q^2\left({(3+2\epsilon)(1+\epsilon)\over3}
+{\epsilon(1-2\epsilon)\over3}\tanh^2\zeta\right)
\nonumber\\
&=&{V_0\over3(1+2\epsilon)}
\exp\left[-\sqrt{2\kappa\epsilon\over1+\epsilon}\phi\right]
\left((3+2\epsilon)(1+\epsilon)+\epsilon(1-2\epsilon)
\exp\left[-\sqrt{2\kappa\over\epsilon(1+\epsilon)}\phi\right]\right)
\nonumber\\
&=&{V_0\over3(1+2\epsilon)}
\left((3+2\epsilon)(1+\epsilon)
\exp\left[-\sqrt{2\kappa\epsilon\over1+\epsilon}\,\phi\right]
+\epsilon(1-2\epsilon)
\exp\left[-\sqrt{2\kappa(1+\epsilon)\over\epsilon}\,\phi\right]
\right).
\end{eqnarray}
Assuming $\epsilon\ll1$, the potential becomes very flat for
$\phi\gtrsim\sqrt{\epsilon/\kappa}$.

Now let us consider the scalar-type perturbation.
The equation (\ref{qeqopen}) for $\bq^p$ in this model becomes
\begin{eqnarray*}
  \left[{d^2\over d\zeta^2}-{4A^2\over\sinh^22\zeta}+p^2\right]\bq^p=0.
\end{eqnarray*}
Setting $z=e^{4\zeta}$, this equation reduces to
\begin{eqnarray*}
 \left[{d^2\over dz^2}+{1\over z}{d\over dz}-{A^2\over (z-1)^2}
 +{A^2\over z(z-1)}+{p^2\over16z^2}\right]\bq^p=0.
\end{eqnarray*}
This can be solved in terms of a hypergeometric function. The
solution is
\begin{eqnarray}
  \bbox{\tilde{q}}^p= e^{ip(\zeta+\Delta\eta)} 
(1-e^{4\zeta})^{1+\epsilon}
{}_2F_1\left[1+\epsilon,1+\epsilon+{ip\over2}\,;1+{ip\over2}\,;
  e^{4\zeta}\right].
\label{solution}
\end{eqnarray}
where we have already choosen the boundary condition
$\bbox{\tilde{q}}^p\to e^{ip\eta}$ as $\eta\to-\infty$.  Computing
the behaviour of the background and of $\bbox{\tilde{q}}^p$ as
$\zeta\to0$, we find
\begin{eqnarray}
a&\to& \frac{\sqrt{1+2\epsilon}}{h0}\zeta^{-1-\epsilon}\nonumber\\
\phi'&\to&\frac{\mu}{4}\zeta^{-1}\\
\bbox{\tilde{q}}^p&\to&e^{ip\Delta\eta}
\frac{\Gamma(1+2\epsilon)\Gamma(1+{ip/2})}
{\Gamma(1+\epsilon)\Gamma(1+\epsilon+{ip/ 2})}
(-4\zeta)^{-\epsilon}
\nonumber,
\end{eqnarray}
from where the scalar transfer function is found to be given by
\begin{equation}
  {\cal T}_S^p = -h_0e^{ip\Delta\eta}2^{-2\epsilon-1/2}\sqrt{\kappa 
    \frac{(\epsilon+1)(2\epsilon+1)}{\epsilon}}
  \frac{\Gamma(1+2\epsilon)\Gamma(1+{ip/2})}
  {\Gamma(1+\epsilon)\Gamma(1+\epsilon+{ip/ 2})}\,.
\end{equation}

For the tensor-type perturbation, the equation (\ref{tensoreq}) for
$\bw$ turns out to be exactly the same as the one for $\bq$ in this
model. Therefore one has $\tbw=\bbox{\tilde{q}}$, and thus
\begin{eqnarray}
  \label{bwexact}
 \tbw^p & \to & e^{ip\eta}= e^{ip(\zeta+\Delta\eta)}, 
\quad (\zeta\to -\infty),
\cr
 \tbw^p & \to &  e^{ip\Delta\eta}
 \frac{\Gamma(1+2\epsilon)\Gamma(1+{ip/2})}
 {\Gamma(1+\epsilon)\Gamma(1+\epsilon+{ip/ 2})}
 (-4\zeta)^{-\epsilon},\quad (\zeta\to0). 
\end{eqnarray}
%\begin{eqnarray*}
% \bw^p_0={e^{ip\Delta\eta}\Gamma(1+2\epsilon)\Gamma(1+{ip\over2})\over 
%  \Gamma(1+\epsilon)\Gamma(1+\epsilon+{ip\over 2})}\,. 
%\end{eqnarray*}
In particular, the phase $\beta$ introduced in Eq.~(\ref{bwlimit}) of
section \ref{tensorin} is found as
\begin{eqnarray}
  \beta&=&\Delta\eta
+{1\over p}
\arg\left({\Gamma(1+{ip/2})\over\Gamma(1+\epsilon+{ip/2})}\right)
\nonumber\\
&\approx&(1-{\pi^2\over12})\epsilon\quad (\epsilon\ll1).
\end{eqnarray}
The last line agrees with the analysis in section \ref{tensorin} under
the small back reaction approximation.
Finally, the tensor transfer function is given by
\begin{equation}
  {\cal T}_T^p = -h_0e^{ip\Delta\eta}4^{-\epsilon}\kappa 
    \sqrt{2\epsilon+1}
  \frac{\Gamma(1+2\epsilon)\Gamma(1+{ip/2})}
  {(p^2+1)\Gamma(1+\epsilon)\Gamma(1+\epsilon+{ip/ 2})}\,.
\end{equation}

\section{Exactly soluble exponential HT model}
\label{HTsoluble}

In \cite{jaumeHT} an analytically soluble exponential model in the
context of the Hawking-Turok Open inflation was considered. In this
model the potential is given by\footnote{In fact, this potential is a
  particular case of the potential discussed in appendix C
  with $\epsilon=1/2$. There only the evolution of the modes after
nucleation is considered, whereas here we consider the normalization
of the primordial spectrum.}
\begin{equation}
V = \frac{3 H_0^2}{\kappa} e^{\sqrt{2\kappa/3}\,\phi}.
\end{equation}
The exact analytical solutions for the background are \cite{jaumeHT}
\begin{eqnarray}
a(\eta_C) = \frac{\sqrt 2}{H_0}\frac{\tanh^{1/2}\eta_C}{\cosh\eta_C}\\
\phi(\eta_C) = -\sqrt{\frac{3}{2\kappa}}\ln\tanh\eta_C,
\end{eqnarray}
where $\eta_C=-\ln(\tan\tau/2)$. The range of $\eta_C$ is
$(0,\infty)$.  In this model the reflection coefficient $\varrho_{+}$
can be analytically computed for both scalar and tensor perturbations,
so a closed expression for the power spectrum can be given (notice
that the transmission coefficient $\sigma_-$ vanishes because
the singularity acts as a reflecting boundary).

We should mention that perturbations in a Hawking-Turok model with 
linear potential has been considered in \cite{cohn}.

\subsection{Scalar power spectrum}
In order to compute the power spectrum of the scalar perturbation on
the comoving hypersurface ${\cal R}_c$, we follow the formalism
described in section \ref{specscal}. We need to find the reflection
coefficient $\varrho_+$ and the special solution
$\bbox{\tilde{q}}^p$ for this model.

To compute $\varrho_+$ in Eq. (\ref{bfqp}), we need to solve the
equation for the perturbations outside the light-cone.  The equation
(\ref{qeq}) for ${\bbox q}$ turns out to be
\begin{equation}
  -{\bbox q}^p{}'' + \frac{3}{\sinh^2 2 \eta_C}{\bbox q}^p 
    = p^2 {\bbox q}^p.
\label{eqscalar}
\end{equation}
The general solution of this equation can be expressed in terms of a
hypergeometric function (just taking Eq.~(\ref{solution}) with
$\epsilon=1/2$ and $\zeta=-\eta_C$),
\begin{equation}
  q^p = e^{-ip\eta_C}(1-e^{-4\eta_C})^{3/2}
  {}_2F_1\left[\frac{3}{2},\frac{3+ip}{2};1+\frac{ip}{2};e^{-4\eta_C}\right].
\end{equation}
and its complex conjugate. This particular solution tends to
$e^{-ip\eta_C}$ as $\eta_C\to \infty$ and diverges as $\eta_C^{1/2}$
when $\eta_C\to0^+$.  Then, the normalized solution which corresponds
to a wave coming from $\eta_C=\infty$ which bounces against the
singularity at $\eta_C\to0^+$ is
\begin{equation}
\label{normsol}
  i{\bbox q}^p_+ = 
  q^p-\frac{\Gamma(1+ip/2)\Gamma(3/2-ip/2)}
    {\Gamma(1-ip/2)\Gamma(3/2+ip/2)}\,\overline{q^p},
\end{equation}
which is everywhere regular. The reflection coefficient $\varrho_+$ can be
read from $i{\bbox q}^p_+$,
\begin{equation}
\label{reflection}
\varrho_+ =-\frac{\Gamma(1+ip/2)\Gamma(3/2-ip/2)}
    {\Gamma(1-ip/2)\Gamma(3/2+ip/2)}.
\end{equation}
Notice that in this case a complete set of modes is given by ${\bbox
  q}^p_+$ with $p>0$.

The modes are continued inside the light-cone
by means of the analytical continuation $\eta_C=-\eta_R-i\pi/2$. Their
evolution inside the bubble is just then given by the analytical
continuation of Eq.  (\ref{normsol}).  In fact, the analytic
continuation of $q^p$ is just the solution $\bbox{\tilde{q}}^p$ we
need to compute the transfer function
\begin{equation}
\label{tildeq}
\bbox{\tilde{q}}^p = e^{ip\eta_R}(1-e^{4\eta_R})^{3/2}
  {}_2F_1\left[\frac{3}{2},\frac{3+ip}{2};1+\frac{ip}{2};e^{4\eta_R}\right]. 
\end{equation}

\begin{figure}[t]
\centerline{\epsfysize=8cm\epsfbox{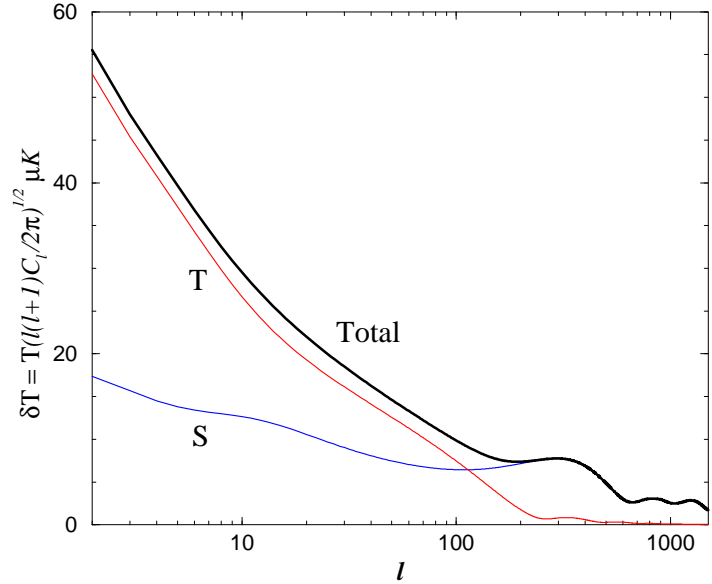}}
\caption{The spectrum of temperature anisotropies
  predicted for the exponential Hawking-Turok model described in
  appendix~\ref{HTsoluble}. We have choosen the
  cosmological parameters $h=0.70$, $\Omega_m=0.35$, $\Omega_B$=0.05,
  $\Omega_\Lambda=0$, $Y_{He}=0.24$, and $N_\nu=3.04$. We show the
  individual contributions from the scalar (S) and tensor (T) modes,
  as well as the total contribution. Since inflation never ends for an
  exponential potential, we have artificially truncated the potential
  so that the number of e-foldings corresponds to a present value of
  the density parameter $\Omega_0=0.4$. Notice that the spectrum is
  finite in spite of the singular nature of the background.}
\label{HTanis}
\end{figure}

Now we compute the assymptotic form of $\bbox{\tilde{q}}^p$ and of
the background as $a_R\to\infty$ . As $\eta_R$ goes to $0$, we find
\begin{eqnarray}
\bbox{\tilde{q}}^p&\to& \frac{1}{\sqrt{\pi}(-\eta_R)^{1/2}} 
\frac{\Gamma(1+ip/2)}{\Gamma(3/2+ip/2)}\,,\\
\phi'&\to&\sqrt{\frac{3}{2\kappa}}\frac{1}{\eta_R}\,,\\
a&\to&\frac{\sqrt 2}{H_0}\frac{1}{(-\eta_R)^{3/2}}\,.
\end{eqnarray} 
Using the results above, the scalar transfer function (\ref{transfersc}) for
this model turns out to be
\begin{equation}
{\cal T}^p_S=H_0\sqrt{\frac{3\kappa}{\pi}}
\frac{\Gamma(1+ip/2)}{(p^2+1)\Gamma(3/2+ip/2)}.
\end{equation}
Using Eq. (\ref{general}), we finally find the primordial power
spectrum for the scalar curvature perturbation,
\begin{equation}
\langle|{\cal R}^p_c|^2\rangle = 
\frac{3\kappa H_0^2}{\pi}\frac{1}{(p^2+4)(p^2+1)}.
\end{equation}
The power spectrum of temperature anisotropies predicted for this
model due to scalar perturbations is shown in Fig.~\ref{HTanis}.

\subsection{Tensor power spectrum}

For the present model, the equation (\ref{weq}) for the tensor perturbation
variable ${\bbox w}^p$  coincides with the one for ${\bbox q}^p$,
\begin{equation}
-{\bbox w}''_p + \frac{3}{\sinh^2 2 \eta_C}{\bbox w}_p = p^2 {\bbox w}_p.
\label{eqtensors}
\end{equation}
The reflection coefficient in Eq. (\ref{bfqpT}) for the tensor
perturbation is then the one computed in the previos section, Eq.
(\ref{reflection}), and the solution $\tbw^p$ coincides with
$\bbox{\tilde{q}}^p$, Eq. (\ref{tildeq}).

Collecting these results, the tensor transfer function
(\ref{generalT}) for this model is given by
\begin{equation}
{\cal T}_T^p = -\frac{2 \kappa H_0}{\sqrt{2\pi}}\frac{\Gamma(1+ip/2)}
{(p^2+1)\Gamma(3/2+ip/2)}\,,
\end{equation}
and the power spectrum by
\begin{equation}
 \langle|U_{p\ell m}^{(+)}|^2\rangle= 
|\tilde {\cal N}^T_p |^2 \left<\left|\frac{Q^p_T}{a^2}\right|^2\right>
       = \frac{8\kappa H_0^2}{\pi(p^2+1)^2}\,.
\end{equation}
The power spectrum of temperature anisotropies predicted for this
model due to tensor perturbations is shown in Fig.~\ref{HTanis}.

\end{document}